\documentclass[aps,prb,reprint,twocolumn,superscriptaddress,amsmath,amssymb]{revtex4-2}%

\usepackage{graphicx}
\usepackage{dcolumn}
\usepackage{bm}
\usepackage[colorlinks,bookmarks=false,citecolor=blue,linkcolor=red,urlcolor=blue,hyperfootnotes=true]{hyperref}

\usepackage{nicefrac}

\begin{document}

\title{The $S=1$ dimer system K$_2$Ni(MoO$_4$)$_2$: a candidate for magnon Bose-Einstein condensation}

\author{B. Lenz}
\email{benjamin.lenz@sorbonne-universite.fr}
\affiliation{IMPMC, Sorbonne Universit\'e, CNRS, MNHN, 4 place Jussieu, 75005 Paris, France.}

\author{B.\ Koteswararao}
\affiliation{Department of Physics, Indian Institute of Technology Tirupati, Tirupati 517506, India}

\author{S.\ Biermann}
\affiliation{CPHT, CNRS, Ecole Polytechnique, IP Paris, F-91128 Palaiseau, France}
\affiliation{Coll{\`e}ge de France, 11 place Marcelin Berthelot, 75005 Paris, France}
\affiliation{Department  of  Physics,  Division  of  Mathematical  Physics, Lund  University,  Professorsgatan  1,  22363  Lund,  Sweden}

\author{P.\ Khuntia}
\affiliation{Department of Physics, Indian Institut of Technology Madras, Chennai  600036, India}
\affiliation{Max-Planck Institute for Chemical Physics of Solids, 01187 Dresden, Germany}

\author{M.\ Baenitz}
\affiliation{Max-Planck Institute for Chemical Physics of Solids, 01187 Dresden, Germany}

\author{S.\ K.\ Panda}
\email{swarup.panda@bennett.edu.in}
\affiliation{Department of Physics, Bennett University, Greater Noida 201310, Uttar Pradesh, India}

\date{\today}

\begin{abstract}
Dimerized quantum magnets provide a unique possibility to investigate Bose-Einstein condensation of magnetic excitations in crystalline systems at low temperature. 
Here, we model the low-temperature magnetic properties of the recently synthesized spin $S=1$ dimer system K${}_2$Ni(MoO${}_4$)$_2$\ and propose it as a new candidate material for triplon and quintuplon condensation. 
Based on a first principles analysis of its electronic structure, we derive an effective spin-dimer model that we first solve within a mean-field  approximation to refine its parameters in comparison to experiment.
Finally, the model is solved by employing a numerically exact quantum Monte Carlo technique which leads to magnetic properties in good agreement with experimental magnetization and thermodynamic results.
We discuss the emergent spin model of K${}_2$Ni(MoO${}_4$)$_2$ in view of condensation of magnetic excitations in a broad parameter regime.
Finally, we comment on a geometrical peculiarity of the proposed model and discuss how it could host a supersolid phase upon structural distortions.
\end{abstract}

\maketitle

\newcommand{\KNi}{K${}_2$Ni(MoO${}_4$)$_2$}
\newcommand{\UP}{$\uparrow$}
\newcommand{\DN}{$\downarrow$}
\newcommand{\idop}{\bm{1}}

\graphicspath{{figures/}}

\textit{Introduction.}
Low-dimensional quantum magnets provide a rich platform to study interesting magnetic phenomena in condensed matter physics due to their inherent strong quantum fluctuations.
A variety of unusual ground states can be realized that sensitively depend on various parameters including dimensionality ($D$), magnitude of the spin ($S$), type of magnetic coupling, or range of correlations, just to name a few. 
Quantum materials thereby offer an ideal alternative route to investigate exotic phases of matter.
A few prominent examples are superfluid and supersolid phases in Bose-Einstein condensates (BEC) \cite{Laflorencie2007,Sengupta2007}, which are usually investigated under extreme conditions in ultracold atoms \cite{Greiner2002} or solid helium-4 \cite{kim2004a,kim2004b}.

In this regard, dimerized quantum magnets have sparked particular interest in recent years due to their inherent BEC of magnetic excitations \cite{Giamarchi2008,Zapf2014}.
These magnets offer the opportunity to study an effective gas of interacting bosons, whose particle number can be tuned by applying an external magnetic field:
$S$ = $\nicefrac{1}{2}$ spin dimers with antiferromagnetic (AF) exchange coupling have a singlet ground state with a finite spin gap to its first excited state of spin $S=1$.
This state, however, becomes the ground state when applying a sufficiently strong external magnetic field - the magnetic moments order and an XY-antiferromagnetic phase is realized.
By mapping the $S=1/2$ spins to hard-core bosons \cite{Matsubara1956,Batista2004}, it turns out that the bosons can condense at this phase transition if the spin environment shows uniaxial symmetry \cite{Batyev84,Giamarchi2008}.
For quantum magnets which fulfill this symmetry condition to a good approximation, the transition from a quantum paramagnetic to a XY-ordered state under an external magnetic field belongs to the BEC universality class.

The ground state properties and their excitations are extensively discussed for a plethora of $S$ = $\frac{1}{2}$  spin dimer materials, which include amongst others 
TlCuCl$_3$ \cite{Oosawa1999}, 
SrCu$_2$(BO$_3$)$_2$ \cite{Kageyama1999},  
BaCu$_2$Si$_2$O$_6$ \cite{Jaime2004}, 
Sr$_3$Cr$_2$O$_8$ \cite{Singh2007} and 
Ba$_3$Cr$_2$O$_8$ \cite{Nakajima2006}. 
Many of the cited spin gap systems exhibit BEC-like excitations under applied magnetic fields or pressure \cite{Zapf2014,Nikuni2000,Ruegg2003,Jaime2004,Aczel2009}. 
On the other hand, very few materials with $S$ = 1 dimers exist in the literature \cite{Uchida2001,Hosokoshi1999}, a famous example being Ba$_3$Mn$_2$O$_8$ \cite{Uchida2001,Uchida2002,Tsuji2005,Stone2008,Samulon2008,Samulon2009,Samulon2010}.
Interestingly, these systems show both triplet and quintuplet excitations:
Whereas $S$ = $\nicefrac{1}{2}$ spin dimer systems exhibit only triplet excitations, a second condensation into the $\vert S=2, S^z=2\rangle$ state is possible for $S=1$ dimer systems in strong magnetic fields.
Since the BEC properties heavily depend on dimensionality, lattice geometry, amount of disorder and the nature of spin interactions \cite{Zapf2014}, new $S=1$ dimer quantum magnets are sought-after to investigate quintuplon condensation.
\begin{figure}
 \includegraphics[width=0.99\columnwidth]{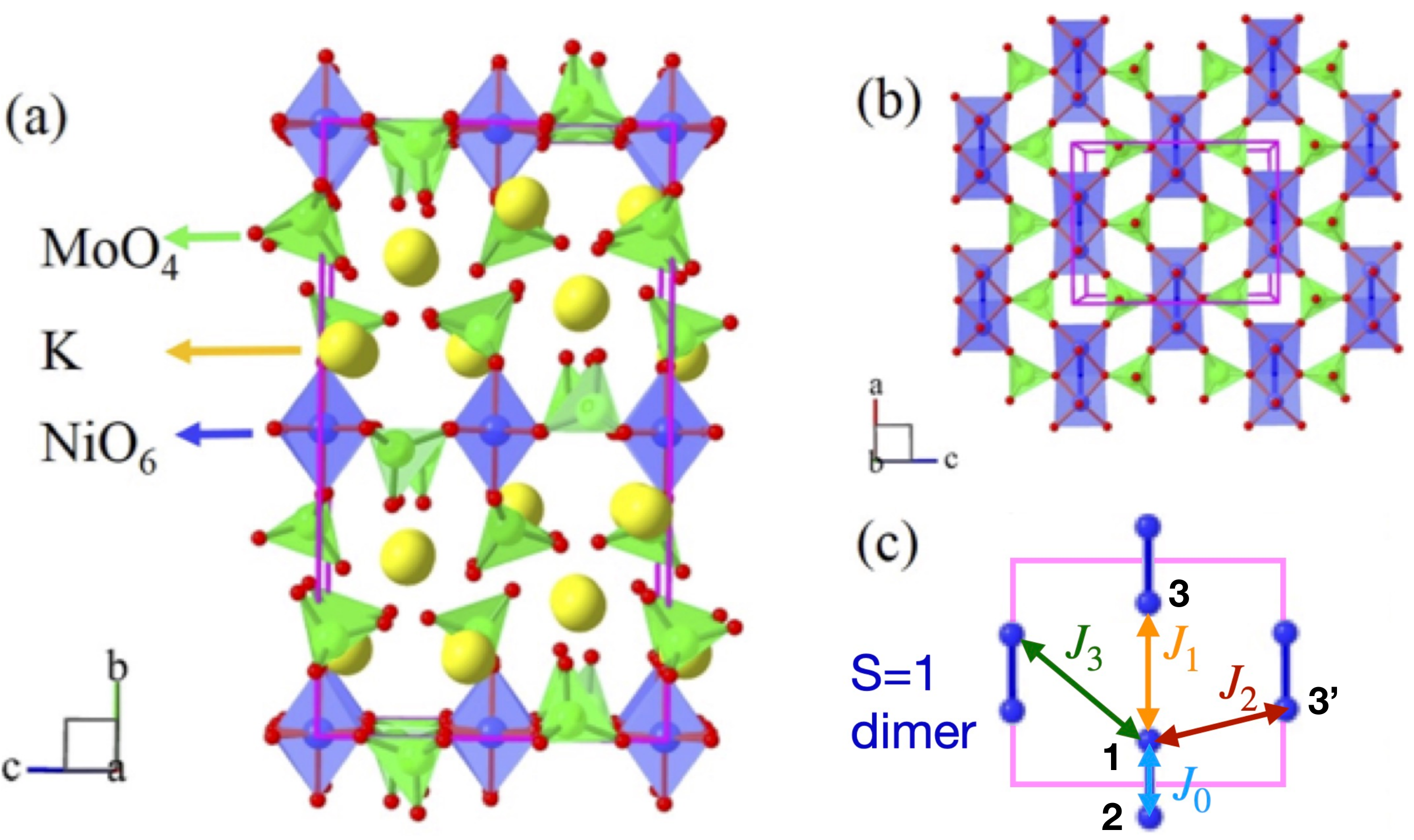}
 \caption{(a) Crystal structure of \KNi. The structure is composed of well-separated 2D layers (\textit{ac}-planes) of Ni$^{2+}$ ($S = 1$) atoms. (b) The \textit{ac}-plane is comprised of edge-sharing NiO$_6$ octahedra, which are connected via MoO$_4$ tetrahedra. Panel (c) shows the arrangements of $S = 1$ dimers in the 2D \textit{ac}-plane, the labels indicate different Ni-Ni spin exchange terms referred to in the text.
 }
 \label{fig:CrystStruct}
\end{figure}

\par
In this regard the recently rediscovered $S$ = 1 spin dimer system K$_2$Ni(MoO$_4$)$_2$ \cite{Murugan2021,Klevtsova1978} is promising:
It has well separated 2D layers (\textit{ac}-planes) that consist of weakly coupled dimers formed by the magnetic ions of Ni$^{2+}$. 
The magnetic susceptibility and heat capacity results \cite{Murugan2021} indicate the presence of a spin gap in the ground state and the magnetization shows a $m=\nicefrac{1}{2}$ plateau characteristic of spin-1 dimer systems.
\par
In this Letter, we derive an effective magnetic model for \KNi\ from first principles calculations and refine it by comparing the computed magnetic properties with the experimental  magnetization results. 
By estimating the inter-dimer spin exchange and mapping onto bosonic excitations, our theoretical modeling predicts superfluid phases of triplons and quintuplons in \KNi.
Moreover, we discuss the possibility of a supersolid phase upon structural distortions assuming a specific parametrization of the inter-dimer spin exchange. 
Our study motivates a more precise determination of the $g$-tensor and spin exchange constants via ESR and neutron scattering experiments in future and suggests an investigation of \KNi\ in the context of both triplon and quintuplon condensation.\\

\begin{figure}[tb]
 \includegraphics[width=0.99\linewidth]{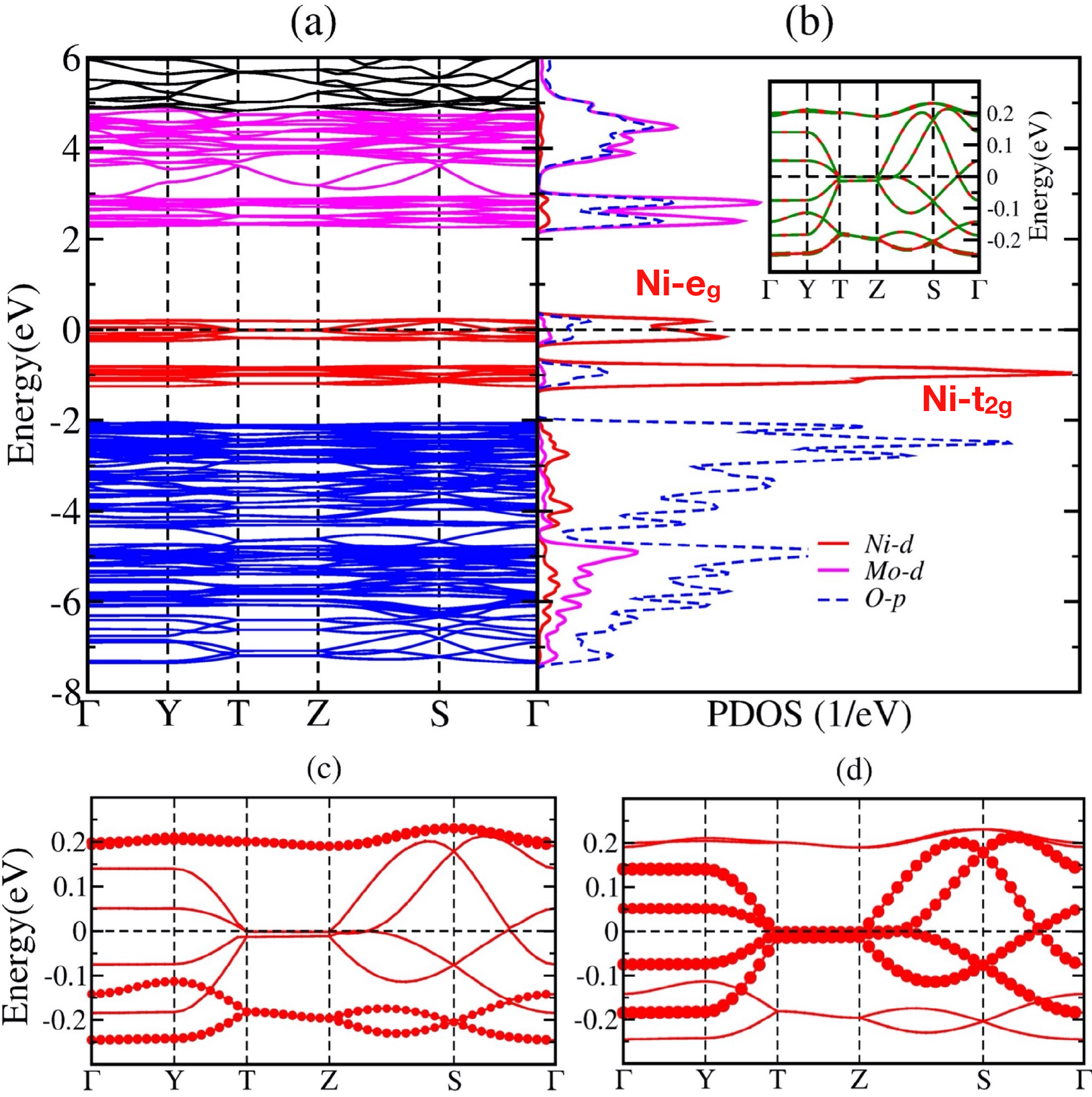}
 \caption{(Color online) (a) The band dispersion along various high symmetry directions within LDA, (b) partial density of states of Ni-$d$, Mo-$d$ and O-$p$ states for non-spin polarized \KNi. Fatband representation of (c) Ni-$x^2-y^2$ and (d) $yz$ orbital character. The inset of panel (b) shows the Wannier-interpolated bands superimposed on the LDA bands. All orbitals are represented in the global reference frame.
}
\label{fig:BandStructure}
\end{figure}

\textit{Model Derivation from first principles.}
In order to interpret the experimental results and to provide a microscopic understanding, we begin our theoretical analysis by carrying out \textit{ab initio} simulations that allow us to derive an effective spin-model.\\
First, we employ (non-spin-polarized) density functional theory (DFT) calculations~\cite{DFT1,DFT2} in the local density approximation (LDA) for the experimentally determined crystal structure. 
As can be seen from the band structure and densities of states (DOS) in Fig.~\ref{fig:BandStructure}(a,b) we find that the $t_{2g}$ states are completely filled whereas $e_g$ states are half filled, as expected for Ni ions in a 2+ charge state (a nominal $d^8$ configuration).
Our analysis shows that these bands around the Fermi level have predominant $d_{x^2-y^2}$ and $d_{yz}$ character in the global reference frame, see Fig.~\ref{fig:BandStructure}(c,d).
By constructing maximally-localized Wannier functions \cite{Wannier} for these bands (see inset of Fig.~\ref{fig:BandStructure}(b)) we obtain a low-energy tight-binding model.
Effective hopping strengths between $e_g$ orbitals (see Table~I of the supplemental information \cite{SupMat}) indicate a strong dimer formation with much weaker inter-dimer coupling.
In particular, the dimerization (e.g. between Ni atoms 1 and 2 in Fig.~\ref{fig:CrystStruct}c) takes place between the $x^2-y^2$ orbitals, in agreement with the pronounced bonding/ anti-bonding splitting of the band structure in Fig.~\ref{fig:BandStructure}c.
This is consistent with fits of the magnetic susceptibility data of Ref.~\onlinecite{Murugan2021} \footnote{Details of the fit and our mean-field estimate of the inter-dimer coupling, which is slightly different from Ref.~\onlinecite{Murugan2021}, can be found in the supplemental information \cite{SupMat}}. 
%
These findings also suggest that the band structure of Fig.~\ref{fig:BandStructure} should only be taken as an indication of the relevant electronic orbitals, since in reality \KNi\ lies deep in the Mott phase.
The half-filled $x^2-y^2$- and $yz$-orbitals should thus be considered as localized state rather than band forming. 
Their low-energy physics is well described by an effective Heisenberg $S=1$ pseudo-spin model describing the co-aligned spins of the half-filled Ni ${e_g}$ orbitals.
Its Hamiltonian is given by 
$$
\hat H = \sum_{i \neq j} J_{ij} \vec{S_i} \cdot \vec{S_j} ,
\label{HH}
$$
where the indices $i$ and $j$ span the positions of the intrinsically magnetic ions in  \KNi, i.e., Ni, and negative (positive) $J_{ij}$ denote (anti-)ferromagnetic spin exchange constants. 
Since $J_{ij}$ scales with the hopping terms $t_{ij}$ as $J_{ij}\sim (t_{ij,x^2-y^2}^2 + t_{ij,yz}^2)$ \cite{SupMat}, the Wannierization of the electronic model suggests to limit the inter-site spin exchange to nearest- and next-nearest neighbors in the \textit{ac}-plane, see Fig.~\ref{fig:CrystStruct}(c).  
The Mott insulating limit of the electronic model also provides a clear hierarchy of the exchange constants, namely $J_0\gg J_1\approx J_2$ \cite{SupMat}. 
Further constraints on the inter-dimer exchange constants $J_1,J_2$ and $J_3$ are obtained from a mean-field treatment of an effective pseudo-spin model, which amounts to fitting the linear regions of the measured magnetization curve under applied external magnetic field as discussed below.\\
\par
\begin{figure}[tb]
	\includegraphics[width=\linewidth]{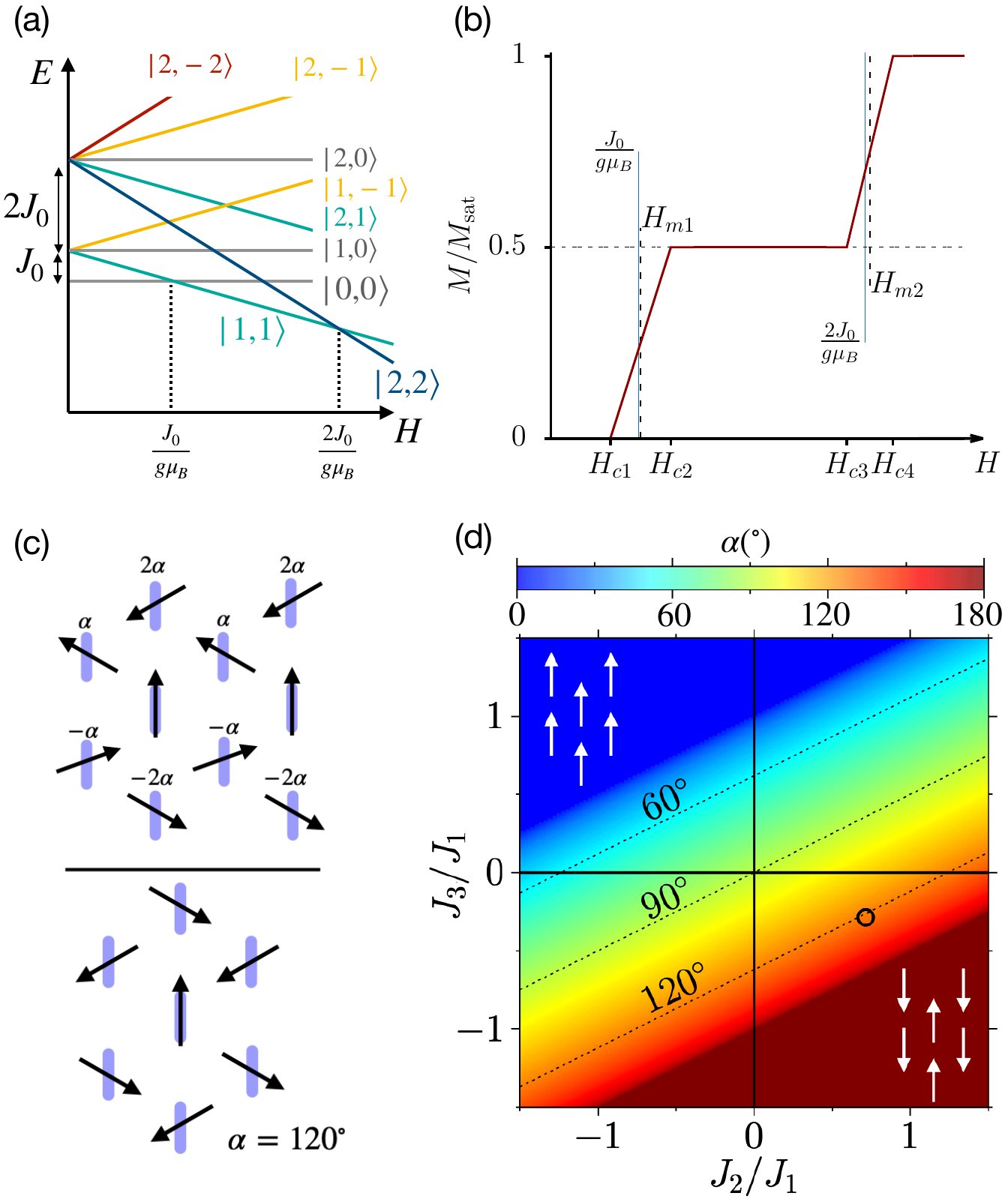}
	\caption{
	(Color online)
		(a) Energy level scheme of an isolated $S=1$ dimer: The groundstate changes from singlet to triplet and finally to quintuplet as the applied magnetic field strength $H$ is increased. 
	    (b) Schematic magnetization curve of the $S=1$ dimer lattice in mean-field theory. Linear slope regions are centered around $H_{m1,2}$.
	    (c) The configurations of the transverse spin components in the magnetization slope regions $\mathcal{I}_1$ and $\mathcal{I}_2$ which are considered here. 
	    $\alpha$ denotes the angle between the y-axis and the transverse pseudo-spin of the dimer, the configuration $\alpha=120^{\circ}$ is close to the parametrization used here, which is indicated with a circle in panel (d).
	    (d) The value of $\alpha$ which minimizes the total energy as a function of $J_2/J_1$ and $J_3/J_1$ for $J_0=33$ K.
	}
	\label{fig:XYconfigs}
\end{figure}
\begin{figure}[tb]
	\includegraphics[width=\linewidth]{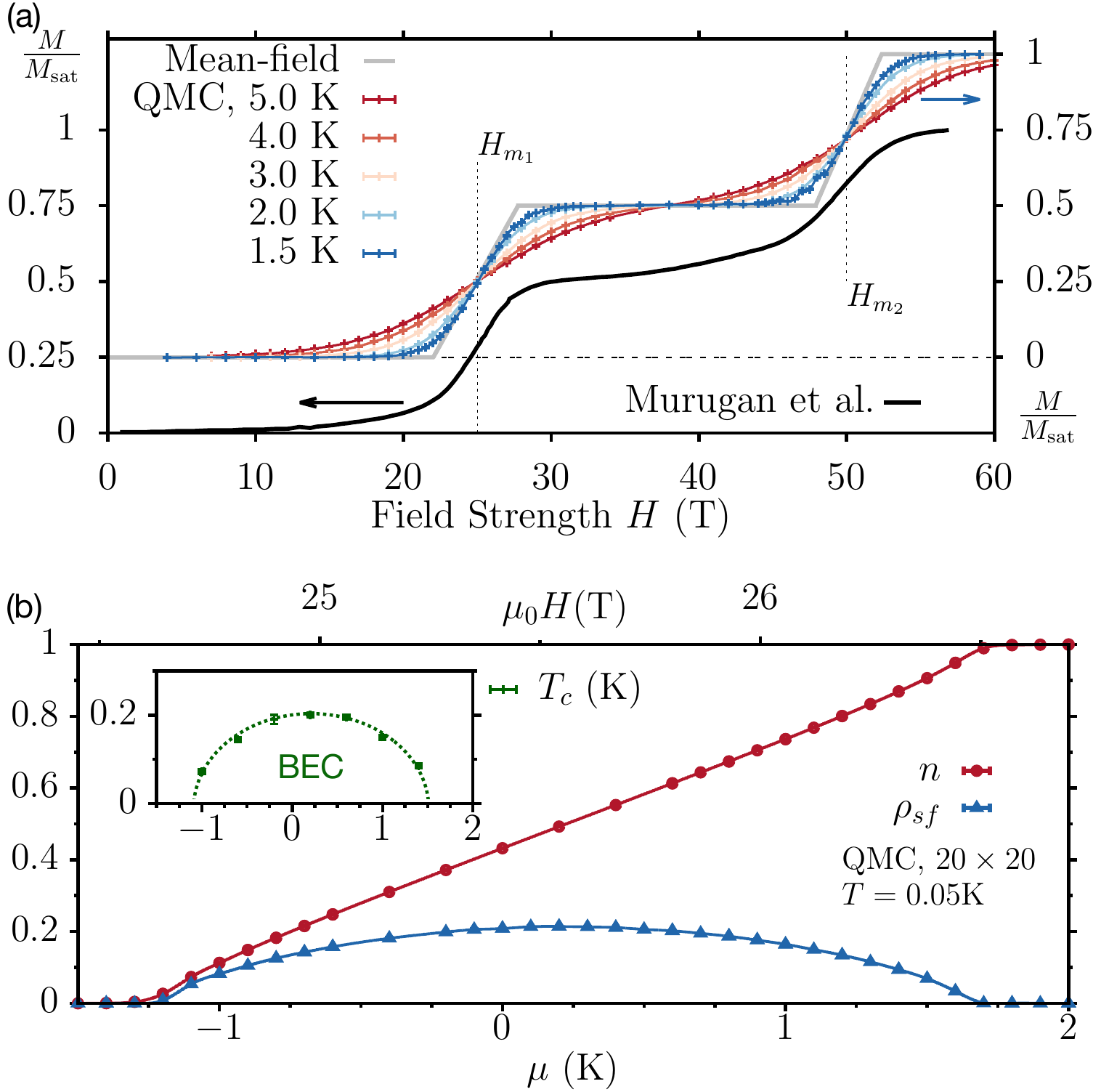}
	\caption{
	    (Color online) 
	    (a) Magnetization as a function of applied magnetic field $H$ as measured at $T=1.5$ K in Ref.~\onlinecite{Murugan2021} and as calculated from QMC simulations according to our model on a $20\times20$ dimer lattice as well as mean-field curve. 
	    {(b) Triplon density $n$ and triplon superfluid stiffness $\rho_{SF}$ of the corresponding hard-core boson model as a function of the chemical potential $\mu$. Inset: Finite-size extrapolated condensation temperature $T_c$. The lines are a guide to the eye.}
	 }
	\label{fig:SpinModel}
\end{figure}

To illustrate this approach, Figure \ref{fig:XYconfigs}(a) sketches the characteristic energy level diagram of an isolated Ni dimer of K$_2$NiMo$_2$O$_8$ as a function of applied magnetic field strength $H$. 
In zero field, the ground state has total spin $S=0$, 
but at $H_1=\frac{J_0}{g\mu_B}$ the ground state changes to a $S=1$ triplet and for $H>H_2=\frac{2J_0}{g\mu_B}$ the dimer is in its $S=2$ quintuplet configuration.
When treating the (weak) inter-dimer interactions in mean-field theory, the three configurations correspond to plateaus in the magnetization curve as sketched in Fig.~\ref{fig:XYconfigs}(b):
A magnetization plateau with half the saturated magnetization is reached for dimers in the triplet state. 
For large magnetic field strength, the dimer is finally in the quintuplet ($\vert2,2\rangle$) state and the magnetization reaches saturation.
Due to the finite inter-dimer exchange terms, the linear magnetization regions around the transition points develop a finite slope.
These regions are characterized by the critical field strengths $H_{c1}$ to $H_{c4}$ and the centers of each linear slope region, $H_{m1},H_{m2}$, can be compared to the critical field strengths extracted experimentally from $dM/dH$. 
In the following, we will briefly revisit key aspects of the mean-field analysis, more technical details can be found in the SI \cite{SupMat}.\\
%
\textit{Mean-Field Calculations.}
Following the technique outlined in the seminal paper by \textit{Uchida et al.} \cite{Uchida2002}, we start by identifying two regions, $\mathcal{I}_1=[H_{c1},H_{c2}]$ and $\mathcal{I}_2=[H_{c3},H_{c4}]$, in which the ground state at \textit{zero temperature} is only composed of the dimer spin states $\vert 0,0\rangle, \vert 1,1\rangle$ and $\vert 1,1\rangle, \vert2,2\rangle$ respectively, see Fig.~\ref{fig:XYconfigs}(a).
In these regions we make the ansatz $\vert \Psi\rangle = \otimes_i \vert \psi_i\rangle$ for the wavefunction, where
\begin{eqnarray}
\vert\psi_i\rangle = 
\begin{cases}
	\cos(\theta_i)\vert0,0\rangle + \sin(\theta_i)\vert 1,1\rangle e^{i\phi_i},\hspace{.3cm}H\in\mathcal{I}_1\\
	\cos(\theta_i)\vert1,1\rangle + \sin(\theta_i)\vert 2,2\rangle e^{i\phi_i},\hspace{.3cm}H\in\mathcal{I}_2.
\end{cases}
\label{Eq:WF}
\end{eqnarray}
In oder to investigate this region further, we rewrite the Hamiltonian in terms of dimer-spin operators and map onto the two lowest-lying states around the critical magnetic field strengths $H_1\ (H_2)$, which allows for a reformulation of the problem in terms of pseudospin-$1/2$ operators $\hat{s}_i$. 
In this new basis, the magnetization of the pseudospins amounts to a change from singlet to triplet (triplet to quintuplet) dimer states around $H_1\ (H_2)$.
Thereby, we obtain a dimer-pseudospin model on a triangular lattice, where the pseudospin magnetization corresponds to the triplet (quintuplet) density.\\
In this description, the phases $\phi_i$ have to be chosen such that they minimize the total energy of the system, which amounts to finding the optimal ordering of the transversal (XY) spin component of the antiferromagnet on a triangular lattice.
To this end, we use the relative phases parametrized by an angle $\alpha$ as sketched in Fig.~\ref{fig:XYconfigs}(c), which leads to different possible relative spin orientations depending on the choice of $J_1,J_2$ and $J_3$, see Fig.~\ref{fig:XYconfigs}(d). \\
Finally, the onsets of the linear slope regions of the magnetization can be expressed as 
\begin{eqnarray}
\begin{split}
g\mu_BH_{c1}&=J_0-\frac{8}{3}b,\ &g\mu_BH_{c3}&=2J_0-2b+a,\\
g\mu_BH_{c2}&=J_0+\frac{8}{3}b+a,\ &g\mu_BH_{c4}&=2J_0+2b+2a,
\label{Eq:Hc}
\end{split}
\end{eqnarray}
where $a=J_1+2J_2+4J_3$ and $b= J_1 + (2J_3-J_2)\cos\alpha - J_1\cos^2\alpha$.	
Based on the measured values of the middle of the linear slope regions, $H_{m1}$ and $H_{m2}$, as well as the critical field strengths $H_{c1},H_{c2}$ determined from a linear fit of the magnetization curve, we estimate the spin exchange constants. 
Since precise information on the $g_{ij}$-tensor is still missing, we assumed a constant $g$-value of $g=2-2.1$.
Future ESR measurements of \KNi\ would allow for a more precise refinement of the model.
In particular the spin exchange $J_3$ depends sensitively on the precise value of $g$, which has consequences for the possibility to host a supersolid phase as discussed below.
In the following, we will use the parametrization obtained for $g=2$, i.e. $J_0=33$ K, $J_1=0.7$ K, $J_2=0.5$ K and $J_3=-0.2$ K.
Treating the inter-dimer spin-exchange in mean-field theory, we find the calculated magnetization curves in good qualitative agreement with the measurements of Ref.~\onlinecite{Murugan2021}, see Fig.~\ref{fig:SpinModel}(a).

\par
To cross check this parametrization, we finally 
perform spin-polarized calculations within the local spin density approximation (LSDA) and LSDA+U (Hubbard $U$)~\cite{LSDAU_FLL}, which assume a static ordering of the spins. 
In both cases a magnetic state corresponding to an anti-parallel spin ordering within and between nearest-neighbor dimers 
is the lowest energy state, consistent with the analysis in Ref.~\onlinecite{Murugan2021}. 
Effective exchange values $J_{ij}$ extracted in a linear-response manner using the magnetic force theorem~\cite{magneticforceth1,magneticforceth2} confirm the parametrization qualitatively \cite{SupMat}. 
%

\textit{Monte-Carlo Results.}
The effective spin model with spin exchange terms $J_0-J_3$ on a triangular dimer lattice can be solved in a numerically exact way in two dimensions using quantum Monte Carlo (QMC) techniques.
To this end, we use the worm QMC algorithm \cite{Prokofev1998,*Prokofev1998b,Troyer2003} as implemented in the ALPS package \cite{ALPS1,*ALPS2}.
For the parameter regime used here ($J_1,J_2$ antiferromagnetic, $J_3$ ferromagnetic, see also Fig.~ \ref{fig:XYconfigs}(d)) there is no fermionic sign problem for the spin lattice, which is why the calculations are rather modest and can be converged with respect to the lattice size:
The results are obtained for $L\times L$ dimer lattices with up to $L=20$ and typically Monte Carlo sampling of $\sim10^{6}$ sweeps with $10\%$ used for thermalization turn out to be sufficient.
\par
Fig.~\ref{fig:SpinModel}(a) shows the evolution of the calculated field-dependent magnetization curve for different temperatures. 
The mean-field result is recovered at low temperature and the curve at $T=1.5$K is in good agreement with the experimental data of Ref.~\onlinecite{Murugan2021}.
To illustrate the BEC of triplons, we plot in Fig.~\ref{fig:SpinModel}(b) the superfluid stiffness $\rho_{SF}$ of the corresponding bosonic model around $H_{c1}<H<H_{c2}$. 
This model is obtained by mapping the triplon excitations onto hard-core bosons, which leads to a spatially anisotropic $t-V$ model on a triangular lattice, see SI \cite{SupMat}.
The triplon density smoothly increases from zero to one triplon per site when tuning the chemical potential across the parameter regime corresponding to the magnetic field strength $H\in\mathcal{I}_1$.
The superfluid stiffness corresponds to the staggered in-plane magnetization $m_{XY}$ of the spin model and indicates condensation of the triplon excitations below the critical temperature $T_c$. 
Since the superfluid density shows -in contrast to the triplon density- considerable finite-size effects, a proper scaling according to the Kosterlitz-Thouless recursion relations is applied \cite{Ceperley1989,SupMat}.
The inset of Fig.~\ref{fig:SpinModel}(b) shows the condensation temperature $T_c$ in the finite-size extrapolated limit, indicating the condensation of triplons around $H_{m_1}$ for $T\lesssim0.2$K.

\textit{Discussion and outlook.}
We note first the qualitative agreement between the calculated magnetization curve and the measurements in Ref.~\onlinecite{Murugan2021}:
The characteristic magnetization plateau at $m=\tfrac{1}{2}$ between $H_{c2}$ and $H_{c3}$ are connected to the zero and saturation magnetization regions at small and high magnetic fields by linear slope regions.
By adjusting the model parameters according to our mean-field analysis, we are able to reproduce the characteristic features of the curve such as the positions of the transitions and the size of the plateau even quantitatively.\\
Differences consist in an early onset of the linear slope region between $13$ T and $22$ T, as well as in a very broad transition from the plateau to saturation magnetization.
Although deviations in high magnetic field might be related to the measurement in high fields, the finite magnetization in smaller magnetic fields $\mu_0 H\sim 20$ T is a robust feature unrelated to uncertainties related to the experimental technique used.
Our QMC simulations at finite temperature do also suggest that these deviations are not finite-temperature effects, since the linear part of the slope $dM/dH$ close to $H_{c}$ is correctly reproduced at $T=1.5$ K. 
Instead, we speculate that these small contributions to the magnetization curve could be linked to contaminations with the related compound ${\mathrm{K}}_{2}{\mathrm{Ni}}_{2}{({\mathrm{MoO}}_{4})}_{3}$ \cite{Koti2017}, which is a spin $S=1$ tetramer system that undergoes a Bose-Einstein condensation at smaller field strength. 
\par
The parametrization of our spin model can also be compared to estimates obtained from fitting the measured susceptibility data.
Since the inter-dimer exchange constants are much smaller than the intra-dimer exchange, one can describe the spin susceptibility to a good approximation with a statistical ensemble of mean-field decoupled spin-1 dimers.
This allows us to extract the intra-dimer exchange constant $J_0$ as well as a mean-field correction due to the inter-dimer exchange, see SI \cite{SupMat}.
Fitting the experimental data after subtracting impurity contributions yields an exchange constant of $J_0=38.8$ K, which is a bit larger than our estimation of $J_0=33$ K.
This is not surprising since the determination of $J_0$ from the susceptibility was shown to deviate from the one via inelastic neutron scattering in similar $S=\nicefrac{1}{2}$ dimer systems by roughly 13\% \cite{Samulon2008,Stone2008}.
In contrast to the analysis carried out in Ref.~\onlinecite{Murugan2021}, we find the mean-field correction $\lambda$ to be finite, $\lambda=\frac{J_1+2J_2-4J_3}{Ng^2\mu_B^2}\approx3$, which is consistent with a small, but finite inter-dimer spin exchange.
However, one should note that the fit is rather insensitive to this quantity, which is why this technique does not allow for a precise determination of the effective inter-dimer exchange \cite{Samulon2008}.
Finally, our estimate of $J_0$ is also in agreement with the spin gap obtained from fitting the magnetic contribution to the specific heat \cite{Murugan2021} ($\Delta\sim38$ K).
\par
The discussed Heisenberg model is the simplest model that qualitatively captures the essential features of the magnetization curve.
A more realistic modeling should also include further terms like single-ion anisotropy, Dzyaloshinskii-Moriya interaction and biquadratic terms. 
However, such a modelization requires a precise knowledge of the different interaction parameters that enter the model and is beyond the scope of this paper.
It could become feasible once ESR and neutron scattering measurements on single crystals allow for determining the inter-dimer interactions with high precision.
It should also be noted that adding a single-ion anisotropy term might change the size of the plateau region, but it would not qualitatively change the shape of the magnetization curve.
In particular, calculations with reasonably-sized single-ion anisotropies did not result in any additional linear slope regions in the magnetization curve that could explain the early onset of a non-zero magnetization found in experiment. 

Finally, we note that the spin-1 Heisenberg model which captures the most prominent features of the system's magnetic properties includes a rather weak inter-dimer exchange term $J_3$, which sensitively depends on the precise value of the Land\'e $g$-factor.
Depending on $g\in[2,2.1]$ either ferro- or antiferromagnetic $J_3$ leads to best agreement with the measured magnetization curves.
In the latter case, the system would be a dimerized spin structure with \textit{frustrated} inter-dimer couplings, which was identified in Ref.~\onlinecite{Sengupta2007} as a crucial criterion for hosting an extended supersolid phase.
\par
Here, however, due to the specific in-plane geometry of the spin-dimers, we did not find supersolid behavior in the effective triplon and quintuplon models \cite{SupMat}. 
The reason lies in the lack of inter-dimer spin frustration along the axis of the dimers.
This is a conceptual difference to the $S=1$ dimer system Ba$_3$Mn$_2$O$_8$, where the dimers are oriented perpendicular to the plane and which in principle allows for such phases. 
\KNi\ thereby not only offers the possibility to investigate Bose-Einstein condensation of triplons and quintuplons as a function of magnetic field, which has so far only been possible in few quantum magnets, but also renders \KNi\ a candidate to tune BEC without supersolid instability.
\par
However, distortions of the crystal structure that lead to either in-plane rotations or out-of-plane buckling of the dimers would naturally induce additional frustrating inter-dimer spin terms that could then allow for a supersolid phase.
The absence of anomalies in the specific heat and magnetic susceptibilities suggest a critical temperature for condensation below $T=1.5$ K, which is confirmed by the derived spin exchange strengths of our modelization.
Overall, our results motivate the investigation of \KNi\ single crystals at low temperature in the future in the context of the realization of emergent states in quantum magnets with exotic magnetic excitations.

\begin{acknowledgments}
B.K. thanks DST INSPIRE faculty award-2014 scheme. 
The figures showing crystal structures were created using the VESTA visualization software \cite{VESTA}.
B.L. acknowledges computation time from TGCC-GENCI (project no. A0110912043) and we thank the CPHT computer team for support.
We thank R. Kumar and A. V. Mahajan for providing additional magnetic measurements.
B.L. thanks Michele Casula for fruitful discussions on QMC and for drawing our attention to Ref.~\onlinecite{Ceperley1989}.
\end{acknowledgments}


%

\clearpage

\section*{Supplemental Material}
\appendix
\setcounter{figure}{0}
\makeatletter
\renewcommand{\thefigure}{A\@arabic\c@figure}
\makeatother

\setcounter{table}{0}
\makeatletter
\renewcommand{\thetable}{A\@Roman\c@table}
\makeatother

\section{Details of the \textit{ab initio} calculations}
\label{sec:AbInitio}
The first step of our \textit{ab initio} analysis consists in performing (non-magnetic) density functional theory (DFT) calculations~\cite{DFT1,DFT2} in the local density approximation (LDA) for the experimentally determined crystal structure. 
After identifying the set of relevant orbitals for the low-energy Hamiltonian, we estimate the hopping strengths ($t_n$) between them by constructing effective maximally-localized Wannier functions for these bands using the WANNIER90 package \cite{wannier90,Wien2wannier}.
The various hopping strengths obtained in this method are given in table~\ref{hopp}.
\bgroup
\def\arraystretch{1.0}
\begin{table}[b]
\caption{Hopping integrals (in meV) between the effective orbitals of nearest- and next-nearest-neighbor Ni ions (indicated in Fig.~1c of the main text and here in Fig.~\ref{fig:sketch}a) as obtained from the Wannierization of our LDA calculation.}
\vspace{0.1cm}
\begin{tabular}{| c c | c c  | c c | c c | }
\hline
 && \multicolumn{2}{c|}{Ni$_2$} & \multicolumn{2}{c|}{Ni$_3$} & \multicolumn{2}{c|}{Ni$_{3^{\prime}}$}\\
  && $yz$ & $x^2-y^2$ & $yz$ & $x^2-y^2$ & $yz$ & $x^2-y^2$  \\
\hline
Ni$_1$& $yz$ & -39 & 0 & 34 & 0 & -38& 4  \\
\hline
Ni$_1$& $x^2-y^2$ & 0 & -201& 0 & 11 & 4 & 18  \\
\hline
\end{tabular}
\label{hopp}
\end{table}
\egroup

The computed $t_n$ clearly reveal a very large effective hopping between the $x^2-y^2$ orbitals of nearest neighbor (NN) Ni ions, indicating a strong dimer formation.
The next nearest-neighbor (NNN) hoppings between the $x^2-y^2$ orbitals are much weaker.
For the $yz$ orbitals, the magnitudes of the hoppings are almost equal for both both NN and NNN and they are almost one order of magnitude smaller than the $x^2-y^2$  NN hopping.
The strong orbital dependence of the hopping parameters can be understood by analyzing the crystal geometry of this system.
The NN $yz$-$yz$ hoppings are primarily mediated via oxygen ions that are shared by the NiO$_6$ octahedra, while the $x^2-y^2$ orbitals of NN nickel ions can hybridize directly via $\sigma$-bonding since Ni ions are positioned on the crystallographic $a$-axis.
As a consequence, the effective NN $x^2-y^2$ hopping  becomes much stronger than the one between $yz$ orbitals.
These results take us to a very important scenario where one of the orbitals renders the system a strong dimer, while the other orbital provides a microscopic root for weaker intra-dimer exchange interactions.
\par
We next determine the lowest energy state by computing energies for a number of possible magnetic states using LSDA and LSDA+U approaches by means of the plane-wave based method as implemented in VASP~\cite{Kresse1996}. 
Our results show that state corresponding to an anti-parallel intra- and inter-dimer spin-alignment is lowest in energy both within LSDA and LSDA+U. 
This  is expected since the $e_g$ states are half filled which promotes antiferromagnetic super-exchange interaction between the Ni-states via the intermediate O atoms. 
The spin moment on the Ni site is calculated to be 1.45~$\mu_B$ within LSDA. 
Inclusion of $U$ helps in localizing the Ni moments and thus increases its value as we increase $U$ within LSDA+U.
The Ni magnetic moment for $U$ = 4 eV and $U$ = 6 eV are respectively, 1.66~$\mu_B$ and 1.77~$\mu_B$.
\par 
After identifying the lowest energy magnetic state, we employed the magnetic force theorem ~\cite{magneticforceth1, magneticforceth2} as implemented in Ref.~\onlinecite{Exchange_Rspt} to estimate the magnetic exchange interactions based on the converged LSDA and LSDA+U simulations. 
The results as summarized in table~\ref{exchange} indicate that the intra-dimer interaction $J_0\approx 40$ K is the dominant one, while the inter-dimer interactions are much smaller. 
Although \textit{ab initio} calculations are not expected to reproduce such small spin-exchange constants with high precision, we note that the obtained value of intra-dimer exchange $J_0$ is in good agreement with the value suggested from the fitting of the experimental magnetic susceptibility data. 
\bgroup
\def\arraystretch{1.2}
\begin{table}[tb]
\caption{Magnetic exchange interactions ($J$) in Kelvin. The various $J$s are marked in Fig.1 of the main manuscript. Positive values indicate antiferromagnetic exchange.}
\centering \vspace{0.1cm}
\begin{tabular}{|c|c|c|c|c|}
\hline
 & $J_0$ & $J_1$ & $J_2$ & $J_3$ \\
 \hline
 LSDA & 154.1 & 5.3 & 2.2 & {-}1.4 \\
 \hline
 LSDA+U ($U$ = 2 eV) & 100.3 & 4.1 & 3.0 & {-}0.4 \\
 \hline
 LSDA+U ($U$ = 4 eV) & 61.8 & 3.0 & 2.3 & {-}0.3 \\
 \hline
 LSDA+U ($U$ = 6 eV) & 38.9 & 2.2 &1.7& {-}0.2  \\
\hline
\end{tabular}
\label{exchange}
\end{table}
\egroup
\par
The magnetic exchange is expected to scale with $\frac{t^2}{U}$, $t$ being the hopping strength and $U$ the correlation strength. 
Thus we clearly see that after inclusion of $U$ within LSDA+U, the magnitudes of all relevant exchange parameters decrease.
Overall, the picture of a system with strong intra-dimer and much weaker inter-dimer exchange is maintained.
However, the relative strengths $J_2/J_1$ and $J_3/J_1$ change qualitatively: Within LSDA+U the nearest-neighbor inter-dimer exchanges are $J_1\sim J_2$, but $J_3\ll -J_1$.
As we show in the following, this hierarchy of exchange constants reproduces well the measured magnetization data.
\par 
Apart from isotropic Heisenberg-exchange interactions, another important term of spin models for BEC in quantum magnets is the magnetic anisotropy which arises from spin-orbit coupling. 
In order to estimate its strength, we carried out LSDA calculations including spin-orbit coupling and fixing the spin-axis to various possible directions. 
The $a$-axis is found to be the easy axis of magnetization with $\sim2$ meV 
lower energy as compared to the $c$-axis, justifying to neglect this term in our simple Heisenberg model.
The orbital moment on the Ni site comes out to be 0.18 ~$\mu_B$.

\section{Derivation of the Effective Spin Exchange}
\label{sec:Exchange}

In the following, we summarize the rigorous derivation of the intra-dimer exchange $J_0$ at zero magnetic field.
Our starting point is the two-orbital dimer with $x^2-y^2$ and $yz$ orbitals separated by a small energy splitting of $\Delta_{e_g}=\epsilon_{d_{yz}}-\epsilon_{d_{x^2-y^2}}$ ($\Delta_{e_g}\approx10$ meV within our DFT simulations) which lifts the degeneracy of the $e_g$ states. 
The dimer is then described by the following Hamiltonian
\begin{equation}
	\mathcal{H} = \sum_{\alpha,\sigma}\left( \hat{c}^{\dagger}_{1,\alpha,\sigma}\hat{c}^{\phantom{\dagger}}_{2,\alpha,\sigma}+ \mathrm{h.c.}\right) + \sum_{i,\alpha}\epsilon_{\alpha}\hat{n}_{i,\alpha} + \mathcal{H}_{\mathrm{int}},
\end{equation}
where $\hat{c}^{(\dagger)}_{i,\alpha,\sigma}$ denotes the annihilation (creation) operator of an electron of spin $\sigma\in\{\uparrow,\downarrow\}$ on orbital $\alpha\in\{d_{yz},d_{x^2-y^2}\}$ at site $i\in\{1,2\}$ and $\hat{n}_{i,\alpha}$ is the density operator of orbital $\alpha$ at site $i$.
For the interaction term we use the Kanamori-Hubbard form
\begin{eqnarray}
\begin{split}
	\mathcal{H}_{\mathrm{int}}=&U\sum_{i,\alpha} n_{i,\alpha,\uparrow}n_{i,\alpha,\downarrow} + U^{\prime}\sum_{i,\alpha} n_{i,\alpha,\uparrow} n_{i,\bar{\alpha},\downarrow} \\
&+ (U^{\prime}-J_H) \sum_{i,\alpha<\alpha^{\prime},\sigma} n_{i,\alpha,\sigma} n_{i,\alpha^{\prime},\sigma}\\
&- J_H\sum_{i,\alpha}c^{\dagger}_{i,\alpha,\uparrow}c^{\phantom{\dagger}}_{i,\alpha,\downarrow}c^{\dagger}_{i,\bar{\alpha},\downarrow}c^{\phantom{\dagger}}_{i,\bar{\alpha},\uparrow}\\
&- J_H \sum_{i,\alpha}c^{\dagger}_{i,\alpha,\uparrow}c^{\phantom{\dagger}}_{i,\bar{\alpha},\uparrow}c^{\dagger}_{i,\alpha,\downarrow}c^{\phantom{\dagger}}_{i,\bar{\alpha},\downarrow}.
\end{split}
\end{eqnarray}
Here, $U$ is the local intra-orbital and $U^{\prime}$ the inter-orbital Hubbard interaction strength, whereas $J_H$ denotes the Hund's coupling strength.
\begin{figure}[tb]
	\includegraphics[width=\linewidth]{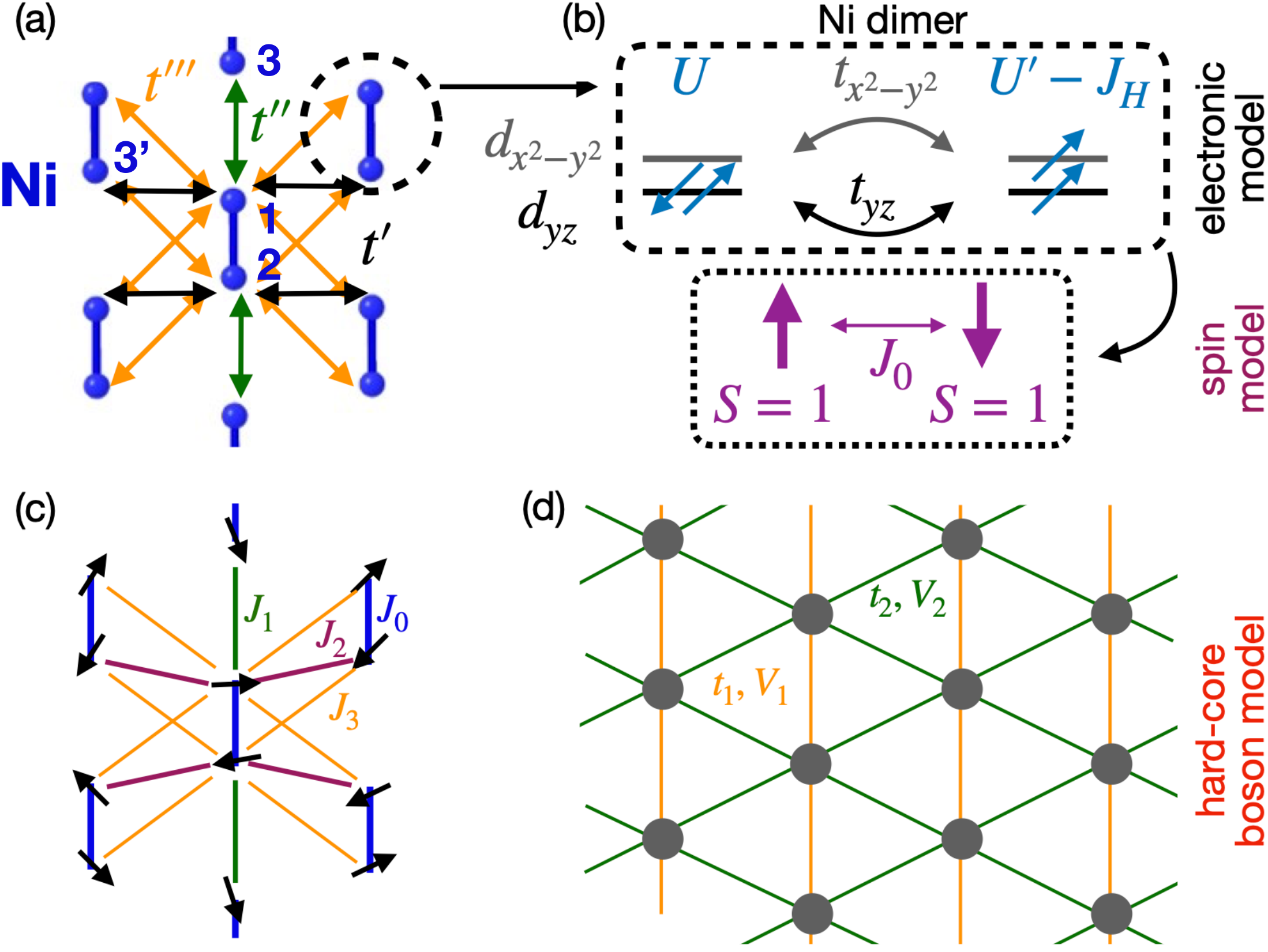}
	\caption{
		(Color online)
		(a) Sketch of the electronic two-orbital model of \KNi\ including inter-dimer hopping terms $t^{\prime},t^{\prime\prime},$ and $t^{\prime\prime\prime}$.
		(b) Each dimer site is mapped to a $S=1$ spin, illustrated on the bottom panel, which gives rise to the spin model shown in panel (c), see also Fig.1(c) of the main text.
		(d) Hard-core boson model from mapping the spin dimers onto its singlet and triplet states for $H\lesssim H_{c1}$.
	}
	\label{fig:sketch}
\end{figure}
\par
We focus on the zero-field behavior and restrict the discussion to the groundstate sector, which is the $\{n_e=4,S^z_{\mathrm{tot}}=0\}$ quantum sector of the dimer.
Furthermore, we consider ony the lowest-order contributions in perturbation theory and neglect higher-order terms.
This means that the excitations considered in the following do not change the orbital occupancy, which is why we choose $\epsilon_0=2\epsilon_{d_{x^2-y^2}}+2\epsilon_{d_{yz}}\equiv 0$ as our reference energy.
In order to map the half-filled two-orbital electronic model to a $S=1$ spin model we use the L\"owdin downfolding technique \cite{Loewdin1951} to derive the effective Heisenberg exchange constants.
\par
Since the Hund's coupling $J_H$ and Hubbard interaction $U$ are the dominating energy scales, the ground state of the system will be in the manifold of six microstates without double occupancy, described by a Hamiltonian $\mathcal{H}_{0,0}$. 
Considering the hopping terms $t_{x^2-y^2}$ and $t_{yz}$ as perturbations, these states are linked to eight excited states, whose Hamiltonian we denote by $\mathcal{H}_{1,1}$. 

The Hamiltonian is then given by 
\begin{equation*}
	\mathcal{H}=\left(\begin{array}{c c}\mathcal{H}_{0,0}&\mathcal{H}_{0,1}\\\mathcal{H}_{1,0}&\mathcal{H}_{1,1}\end{array}\right),
\end{equation*} 
where $\mathcal{H}_{0,1}=\mathcal{H}_{1,0}^{\dagger}$ include the hopping terms $t_{x^2-y^2}$ and $t_{yz}$.

L\"owdin downfolding onto the subspace of zero double occupation leads in lowest order to the approximate Hamiltonian
\begin{equation}
	\tilde{\mathcal{H}}(\epsilon)=\mathcal{H}_{0,0} + \frac{\mathcal{H}_{0,1}\mathcal{H}_{1,0}}{\epsilon - U - 2U^{\prime}}.
	\label{eq:Hdownfold}
\end{equation} 

Next, we express the dimer spin-1 states $\vert J,M_J\rangle = \vert 0,0\rangle,\ \vert1,0\rangle,$ and $\vert2,0\rangle$ in the electronic basis, and compare their energies to the downfolded Hamiltonian.
From the $S=1$ dimer picture we expect at zero magnetic field a singlet-triplet splitting of $J_0$ and a triplet-quintuplet splitting of $2J_0$, which is consistent with $J_0(\epsilon) = \frac{t_{x^2-y^2}^2+t_{yz}^2}{\epsilon-U-2U^{\prime}}.$

We finally make the static approximation for a characteristic energy of the system.
If we choose $\tilde{\epsilon}=2U^{\prime}-J_H$, which  yields
\begin{equation}
	J_0 = -\frac{t_{x^2-y^2}^2+t_{yz}^2}{U+J_H},
	\label{Eq:J0}
\end{equation}
we recover the energy level scheme of our electronic two-orbital model.


\section{Dimer fit to the susceptibility}
\label{sec:FitSusc}
The inter-dimer magnetic exchange constants $J_1,J_2,J_3$ are much smaller than the intra-dimer exchange $J_0$, which renders a static mean-field decoupling of the dimers valid at sufficiently high temperatures. 
Here, $J_1,J_2,J_3< 2$ K, which is why the susceptibility data suggests itself to determine the intra-dimer exchange $J_0$. 
Therefore, by assuming a statistical ensemble of decoupled spin-1 dimers one obtains the susceptibility \cite{Uchida2001}
$$\chi_0 = \frac{2Ng^2\mu_B^2}{k_BT}\frac{1+5e^{-2J_0/k_BT}}{3+e^{J_0/k_BT}+5e^{-2J_0/k_BT}},$$
where $N$ denotes the number of spin-1 dimers.
Treating the inter-dimer exchange terms in static mean-field theory leads to 
\begin{equation}
\chi = \frac{\chi_0}{1-\lambda\cdot\chi_0},\label{Eq:chi}
\end{equation}
with $\lambda=\frac{J_1+2J_2-4J_3}{Ng^2\mu_B^2}$. 
Most importantly, the temperature $T_{\mathrm{max}}$, where $\chi(T)$ reaches its maximum is also the position of the maximum of $\chi_0(T)$. 
The maximum of $\chi(T)$ ist therefore well suited to determine $J_0$.

\begin{figure}[tb]
	\includegraphics[width=\linewidth]{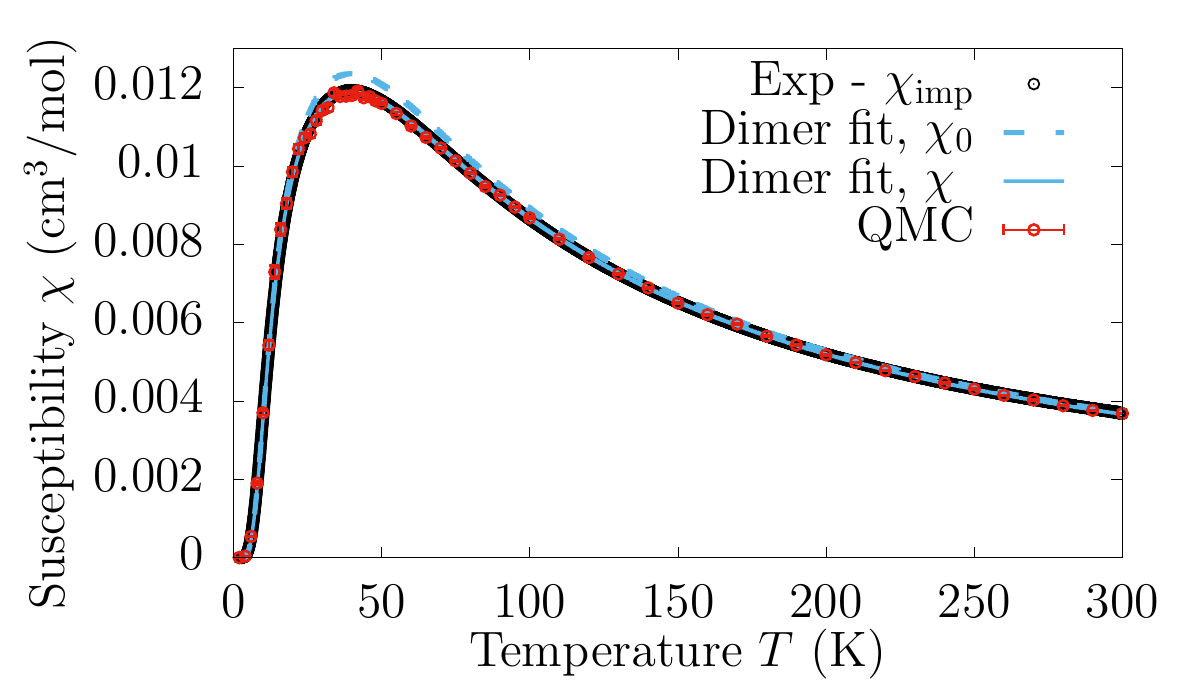}
	\caption{
	    (Color online) 
	    Measured magnetic susceptibility $\chi(T)$ as well as the spin-1 dimer fit to Eq.(\ref{Eq:chi}) and result of the QMC calculation on a $16\times16$ dimer lattice.
	}
	\label{Fig:ChiFit}
\end{figure}

Fitting the measured data after substracting impurity contributions (see main text) leads to $J_0=38.8K$, $\lambda=3.12\pm0.01$ and $2N\mu_B^2g^2=1.83$.
By assuming that the Heisenberg exchange scales as $t_{x^2-y^2}^2+t_{yz}^2$, see section \ref{sec:Exchange}, we note that the parameters $J_1$ and $J_2$ have nearly the same size, which is why we set $J_1=J_2, J_3=0$ for a first simplistic modelization such that $\lambda=3$. 
Using $J_0=38.8$ K and $J_1=J_2=1$ K, the QMC simulation is found to be in perfect agreement with both the experiment and the dimer fit at temperatures $T>2K$, see Fig.~\ref{Fig:ChiFit}.

\section{Details of the mean-field analysis of the magnetization curve}
\label{sec:MFMag}
In the following we will analyze the magnetization curve by treating the inter-dimer spin exchange terms on a mean-field level by closely following the formalism outlined in Ref.~\onlinecite{Uchida2002}.

We start by identifying two regions, $\mathcal{I}_1=[H_{c1},H_{c2}]$ and $\mathcal{I}_2=[H_{c3},H_{c4}]$, in which the ground state at zero temperature is composed of the dimer spin states $\vert S=0,S^z=0\rangle, \vert 1,1\rangle$ and $\vert 1,1\rangle, \vert2,2\rangle$ respectively.
In this regions we make the ansatz $\vert \Psi\rangle = \otimes_i \vert \psi_i\rangle$ for the wavefunction, where
\begin{equation}
\vert\psi_i\rangle = 
\begin{cases}
	\cos(\theta_i)\vert0,0\rangle + \sin(\theta_i)\vert 1,1\rangle e^{i\phi_i},\hspace{.5cm}H\in\mathcal{I}_1\\
	\cos(\theta_i)\vert1,1\rangle + \sin(\theta_i)\vert 2,2\rangle e^{i\phi_i},\hspace{.5cm}H\in\mathcal{I}_2.
\end{cases}
\label{Eq:WF}
\end{equation}
Furthermore we make the assumption that $\theta_i=\theta\ \forall i$.
The expectation values of the spin operators acting on sites $1,2$ of dimer $i$ read as
\begin{eqnarray*}
\langle S_{i,1}^z\rangle = \langle S_{i,2}^z\rangle &=& \frac{1}{2}\sin^2(\theta)\\
\langle S_{i,1}^{\pm}\rangle = -\langle S_{i,2}^{\pm}\rangle &=& -\frac{2}{\sqrt{3}} \cos(\theta)\sin(\theta)e^{\mp i \phi_i},
\end{eqnarray*}
for $H\in\mathcal{I}_1$ and 
\begin{eqnarray*}
\langle S_{i,1}^z\rangle = \langle S_{i,2}^z\rangle &=& \frac{1}{2}\left(1+\sin^2(\theta)\right)\\
\langle S_{i,1}^{\pm}\rangle = -\langle S_{i,2}^{\pm}\rangle &=& - \cos(\theta)\sin(\theta)e^{\mp i \phi_i},
\end{eqnarray*}
for $H\in\mathcal{I}_2$.

In this description, the phases $\phi_i$ have to be chosen such that they minimize the total energy of the system, which amounts to finding the optimal ordering of the transversal (XY) spin component of the antiferromagnet on a triangular lattice.
The configuration naturally depends on the values of $J_1,J_2$ and $J_3$.
Here, we use the relative phases sketched in Fig.~4(a) of the main text, which leads to different possible angles $\alpha$ as discussed below.

The corresponding total energies in the two regions read as
\begin{eqnarray*}
E_1 =&& J_0 (\sin^2\theta - 2) - g\mu_BH\sin^2\theta\\
&&+ \frac{a}{2}\sin^4\theta - \frac{8}{3}\cos^2\theta\sin^2\theta \cdot b
\end{eqnarray*}
and 

\begin{eqnarray*}
E_2 =&&J_0 (2\sin^2\theta - 1) - g\mu_BH(1+\sin^2\theta) \\
&&+ \frac{a}{2} \left(1+2\sin^2\theta+\sin^4\theta \right)- 2\cos^2\theta\sin^2\theta\cdot b
\end{eqnarray*}
with $a=J_1+2J_2+4J_3$ and $b= J_1 + (2J_3-J_2)\cos\alpha - J_1\cos^2\alpha$.	

The relative phase angle $\alpha$ is determined from $\frac{\mathrm{d}E}{\mathrm{d}\alpha}=0$, which leads to up to three different solutions:
$$b = \begin{cases}
	-J_1-2J_2+4J_3,\hspace{.5cm}&\alpha=0\\
	-J_1+2J_2-4J_3,\hspace{0.5cm} &\alpha=\pi\\
	\phantom{-}J_1+\frac{(2J_3-J_2)^2}{2J_1},\hspace{.5cm} &\alpha=\arccos\frac{2J_3-J_2}{2J_1},\\&\mathrm{\ if\ } \vert 2J_3-J_2\vert \leq\vert2J_1\vert\\
\end{cases}$$
Depending on the values of $J_2/J_1$ and $J_3/J_1$, the solution which maximizes $b$ (and thereby minimizes $E$) is chosen.
The corresponding angle $\alpha$ is plotted in Fig.~3c) of the main text.
\begin{figure}[tb]
	\includegraphics[width=\linewidth]{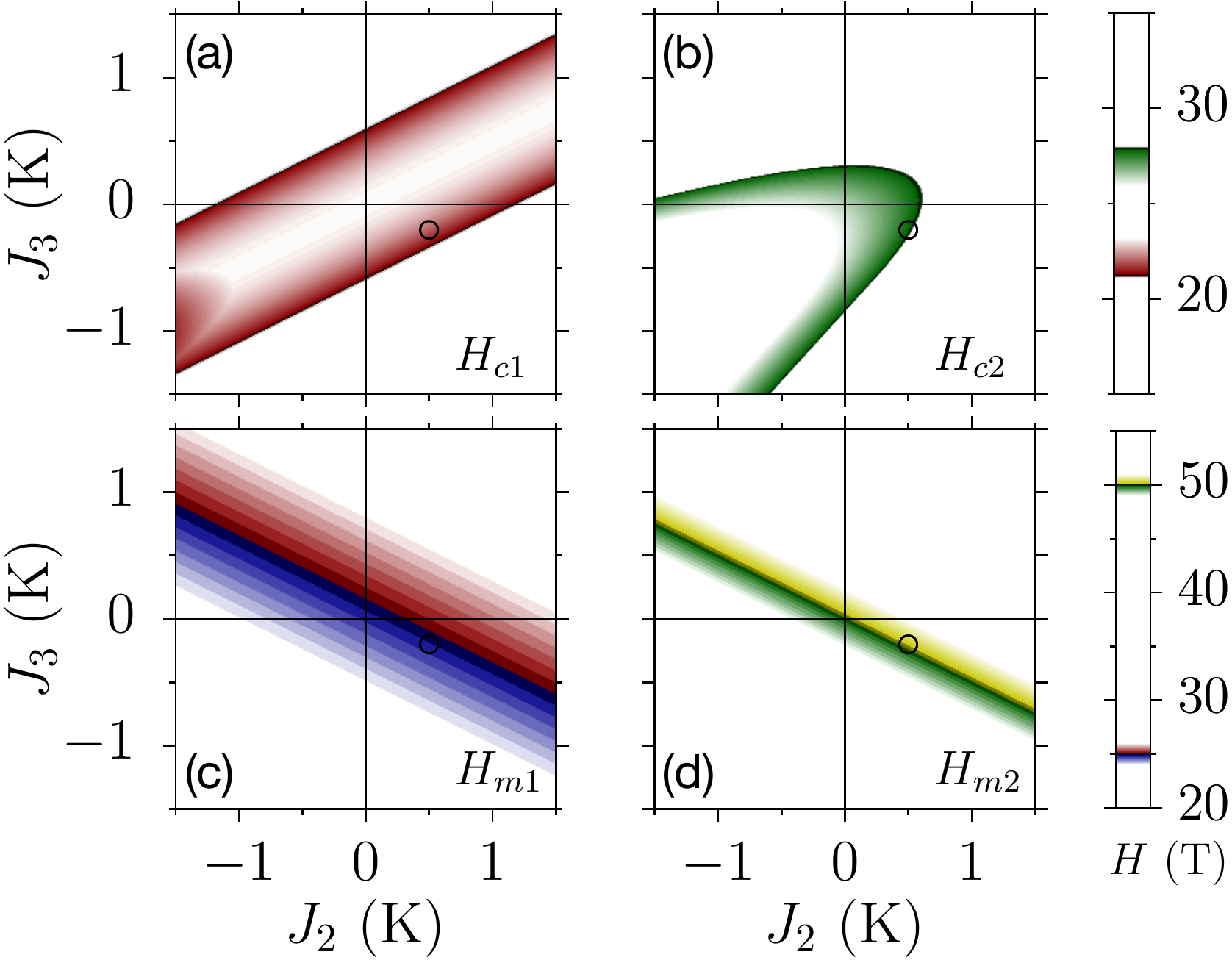}
	\caption{
	(Color online)
	    The critical field strengths $H_{c1}$ (a) and $H_{c2}$ (b) according to Eq.~(\ref{Eq:Hc}) for $J_0=33$ K and $J_1=0.7$ K. 
	    Regions $[\tilde{H}_{c1},\tilde{H}_{c1}2]$ and $[\tilde{H}_{c2}-2,\tilde{H}_{c2}]$ are coloured, where $\tilde{H}_{c1(c2)}$, shown in black, are taken as lower (upper) limits from fits to the measured magnetization curve at $T=1.5$ K.
	    Panels (c) and (d) show the middle of the linear slope regions, $H_{m1,2}$, as a function of $J_2$ and $J_3$.
	    The estimations from Ref.~\cite{Murugan2021} based on the derivative of the magnetization curve are shown in black.
	    The parametrization used in the paper is indicated with a circle.
	}
	\label{fig:MF_Hcs}
\end{figure}
In order to calculate the magnetization $m$, we require $\frac{\mathrm{d}E}{\mathrm{d}\theta}=0$ and solve for $\sin^2\theta$:
$$
\sin^2\theta=
\begin{cases}
	\frac{g\mu_BH-J_0+\frac{8}{3}b}{a+\frac{16}{3}b},\hspace{.5cm}&H\in\mathrm{I}_1\\
	\frac{g\mu_BH-2J_0+2b-a}{a+4b},\hspace{.5cm}&H\in\mathrm{I}_2
\end{cases}
$$
This means that 
\begin{eqnarray}
\begin{split}
g\mu_BH_{c1}&=&J_0-\frac{8}{3}b\\
g\mu_BH_{c2}&=&J_0+\frac{8}{3}b+a\\
g\mu_BH_{c3}&=&2J_0-2b+a\\
g\mu_BH_{c4}&=&2J_0+2b+2a.
\label{Eq:Hc}
\end{split}
\end{eqnarray}
In the regions $\mathcal{I}_1$ and $\mathcal{I}_2$, the slope of the magnetization curve reads as $g_1= \frac{3g\mu_B}{3a+16b}$ and $g_2=\frac{g\mu_B}{a+4b}$, which means that $g_1^{-1}=g_2^{-1}+\frac{4}{3g\mu_B}b$.
Furthermore, the middle of the magnetization slope regions $\mathrm{I}_{1},\mathrm{I}_{2}$ are given by $H_{m1}=J_0+\frac{a}{2}$ and $H_{m2}=2J_0+\frac{3}{2}a$ respectively.
Since these positions barely change at low temperature, see Fig.~4(b) of the main paper, 
we can use the magnetization data measured at $T=1.5$ K to estimate $J_0$.

At $T=1.5$ K the slope of the linear region of the magnetization curve is smaller than the zero-temperature value which should be used for the mean-field approximation.
Therefore, the critical field strengths $\tilde{H}_{c1}$ - $\tilde{H}_{c4}$ determined from a linear fit of the measured data serve as upper bounds to the slope regions $\mathcal{I}_{1}, \mathcal{I}_{2}$.
Since the low-field data is slightly more reliable than the high-field data, we use the linear region at intermediate fields to fit  $\tilde{H}_{c1}=21.2$ T and $\tilde{H}_{c2}=27.9$ T. 
Within the mean-field treatment, the gradients of the two slope regions are connected.
For values of $J_1$ in the order of $\sim 1$ K the parameter $b$ is found to be $b\sim g\mu_B$, which leads to the magnetization slope of the region $\mathcal{I}_2$. 
By determining the middle of these slope regions, we finally estimate $J_0\approx 33$ K, $a\approx 2$ K.

\par
Although a thorough determination of the parameters $J_1,J_2,J_3$ would require neutron scattering data on single crystals, even the present data allows to get further insights from the mean-field solution.
We know from the electronic model discussed in section \ref{sec:Exchange} that the exchange constants scale as $t_{x^2-y^2}^2+t_{yz}^2$.
Together with the LSDA+U analysis of section \ref{sec:AbInitio} we conclude that $J_2\lesssim J_1$.
We fix $J_1=0.7$ K and $g=2$ and determine the remaining parameters $J_2$ and $J_3$ by taking the extracted values $\tilde{H}_{c}$ as upper/lower bounds for $H_{c2,4/c1,3}$, see Fig.~\ref{fig:MF_Hcs}.
Therefore we set $J_2=0.5$ K and $J_3=-0.2$ K, which leads to good agreement of the finite-temperature QMC calculations with experiment as shown in Fig.~3(b) of the main text.

\par
Since ESR measurements of the similar quantum dimer system BaCuSi$_2$O$_6$ showed that the $g$ factor is slightly larger than $2$ \cite{Zvyagin2006}, we briefly investigate the effect of $g>2$ on the optimal values of $J_2$ and $J_3$.
In particular the small inter-dimer exchange $J_3$ is susceptible to small variations in $g$:
For $g=2.07$, the best parametrization would correspond to an \textit{antiferromagnetic} exchange $J_3=0.2$ K, which has consequences for the corresponding hardcore boson model that describes the Bose-Einstein condensation of triplons.
\begin{figure}[tb]
	\includegraphics[width=\linewidth]{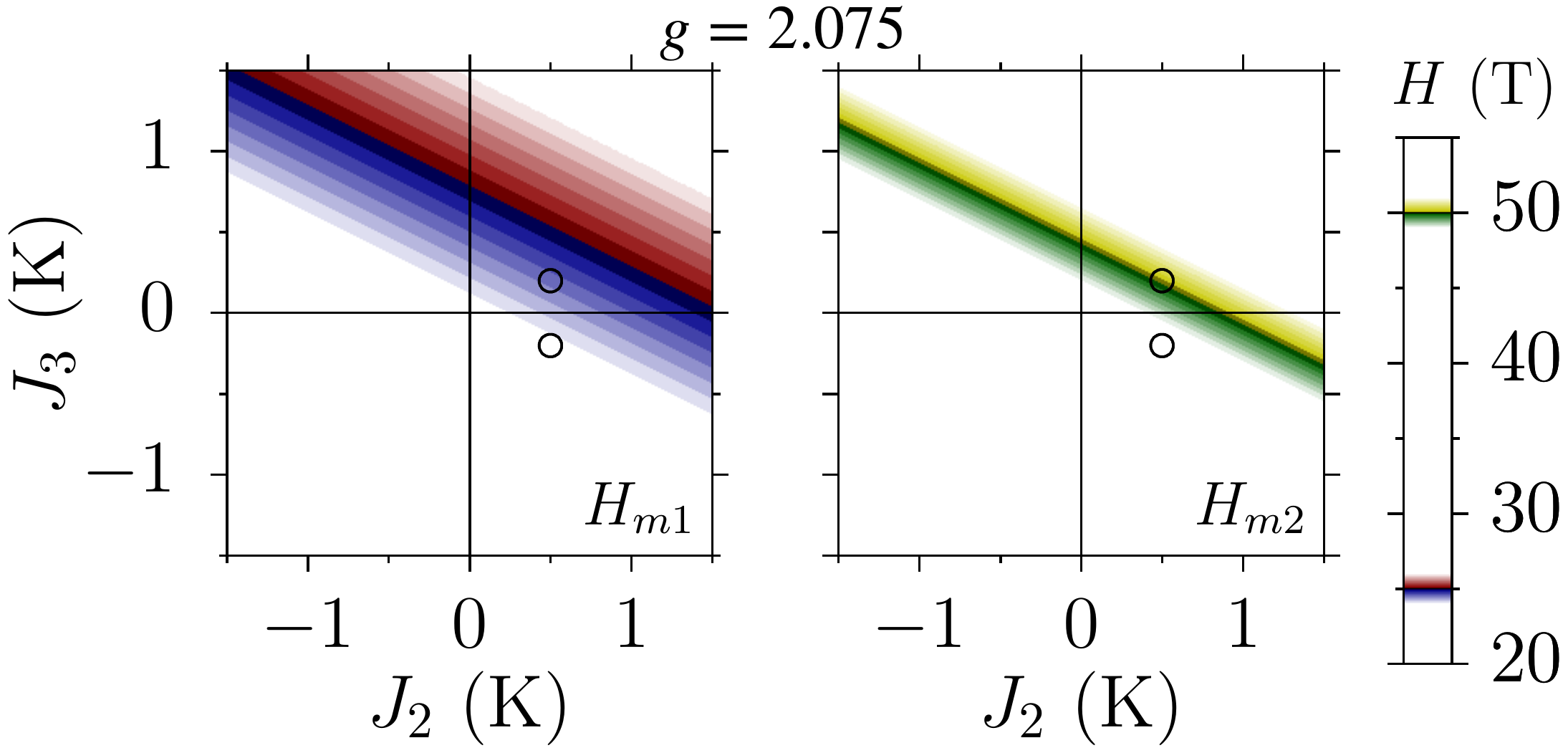}
	\caption{
	(Color online)
	    Same as in Fig.~\ref{fig:MF_Hcs}(c)-(d), but this time for $g=2.075$ instead of $g=2$. For the same values of $J_1$ and $J_2$ one finds an antiferromagnetic spin exchange $J_3=0.2$ K as indicated by the second circle:
	    Depending on the precise value of $g$ one obtains $J_3\in[-0.2\mathrm{K},0.2\mathrm{K}]$. 
	}
	\label{fig:MF_Hcs_g}
\end{figure}
%

\section{Construction of an effective bosonic model}
We focus here on the condensation of triplons and derive a hardcore boson model for magnetic field strengths up to $H\lesssim H_{c2}$.
Note that since we neglect in our simplified spin model anisotropy terms such as single-ion anisotropy and since triplon and quintuplon excitations are well separated, the calculation of quintuplon condenssation would proceed likewise.\\
Following Ref.~\onlinecite{Zapf2014}, we use a Matsubara-Matsuda transformation \cite{Matsubara1956,Batista2004} to derive an effective hardcore boson model:
Each dimer is mapped to a lattice site, which is empty if the dimer is in its singlet configuration and can host up to one boson, which corresponds to the dimer triplet state $\vert 1,1\rangle$.
The emerging bosonic model corresponds to a $t-V$ model on an anisotropic triangular lattice:
\begin{eqnarray*}
 \mathcal{H} =& -t_1\sum_{\langle i,j\rangle} (b^{\dagger}_ib^{\phantom{\dagger}}_j + h.c.) -t_2\sum_{\langle\langle i,j\rangle\rangle} (b^{\dagger}_ib^{\phantom{\dagger}}_j + h.c.) \\
 &+ V_1\sum_{\langle i,j\rangle} n_i n_j + V_2\sum_{\langle\langle i,j\rangle\rangle} n_i n_j - \mu\sum_i n_i,
 \end{eqnarray*}
where $t_{1,2}$ denote bosonic hopping along the lattice vectors $\vec{a}_1,\vec{a}_2$ as depicted in Fig.~\ref{fig:sketch}(d) and $V_{1,2}$ are the inter-site interaction strengths.
This corresponds to a square lattice with additional hopping and interaction terms along one of the diagonals.
Changing the magnetic field strength $H$ translates into modifying the chemical potential $\mu=-J_0+g\mu_B H - \frac{1}{2}\sum_r V_{r,r^{\prime}}$.
For the (isotropic) hardcore boson $t-V$ model, a superfluid phase is realized for $t>V$, but for $t\ll V$ a supersolid phase exists \cite{Boninsegni2005,Wessel2005,Heidarian2005,Melko2005}. 
In the spatially anisotropic triangular lattice, the supersolid phase can also be found in a certain parameter regime \cite{Gan2008} and even an incommensurate supersolid phase can be stabilized \cite{Zhang2016}.
Here, however, the geometry of the spin-dimer lattice causes a slightly different scenario, which we discuss in the following.\\
Since only $J_1$ couples dimers along the $\vec{a}_1$ direction, we obtain $t_1=\frac{J_1}{4}=V_1$.
For the other directions the model parameters read
$$t_2=\frac{J_2-2J_3}{4},\ \ V_2=\frac{J_2+2J_3}{4}.$$
Depending on the precise choice of $J_3\in[-0.2\mathrm{K},0.2\mathrm{K}]$, the ratio $t_2/V_2$ changes from
$t_2> V_2$ for ferromagnetic $J_3$ to $t_2< V_2$ for antiferromagnetic (frustrated) $J_3$.

\section{Bose-Einstein condensation in the effective model}
\begin{figure}[tb]
	\includegraphics[width=\linewidth]{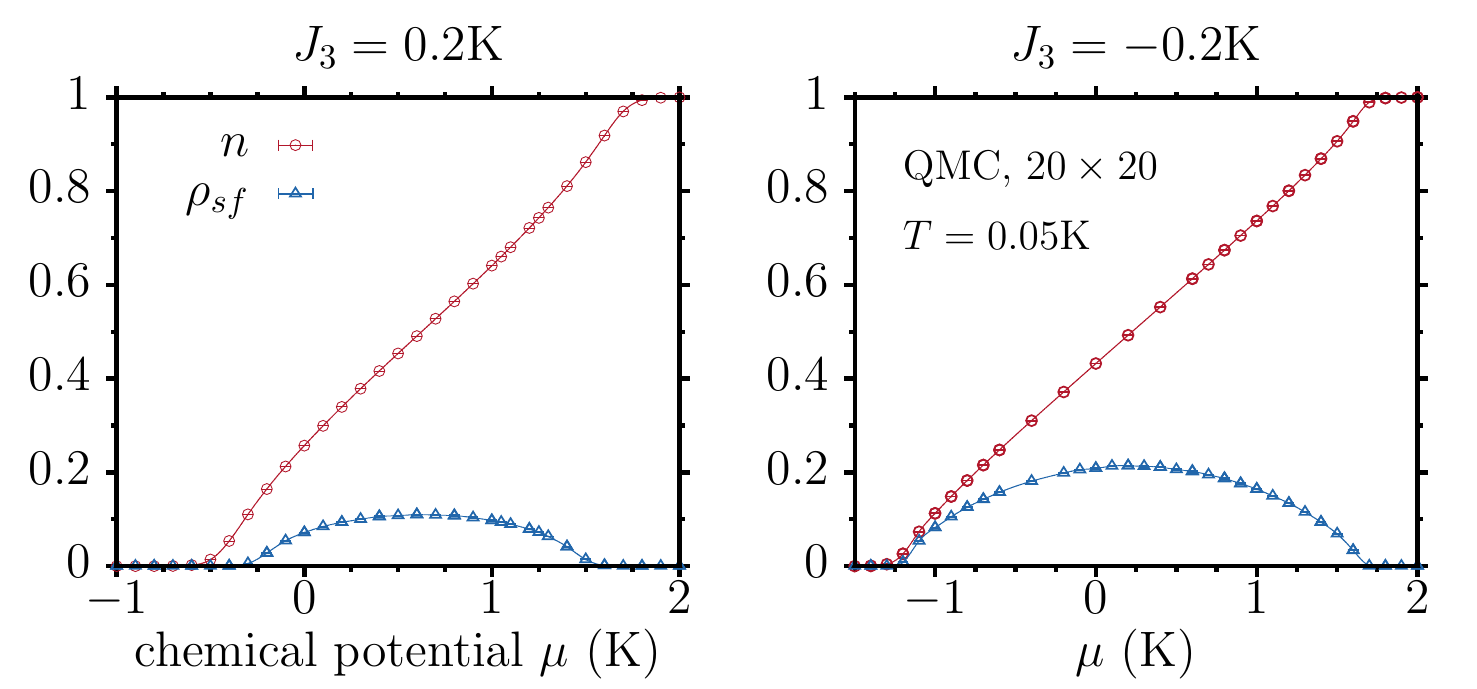}
	\caption{
	(Color online)
	    Density of hard-core bosons, $n$, and superfluid stiffness $\rho_{SF}$ as a function of chemical potential $\mu$ for the effective hard-core boson model discussed in the text.
	     The parameter region shown corresponds to magnetic field strengths $H_{c1}\lesssim H\lesssim H_{c2}$ of the spin model.
	     We show results for the parametrization used in the main paper ($J_3=-0.2$ K, right panel) as well as for the alternative parametrization with an antiferromagnetic spin exchange ($J=0.2$ K, left panel), which is found for $g=2.075$. These parametrizations are indicated by circles in Fig.~\ref{fig:MF_Hcs_g}.
	}
	\label{fig:BEC}
\end{figure}

\begin{figure}[tb!]
	\includegraphics[width=\linewidth]{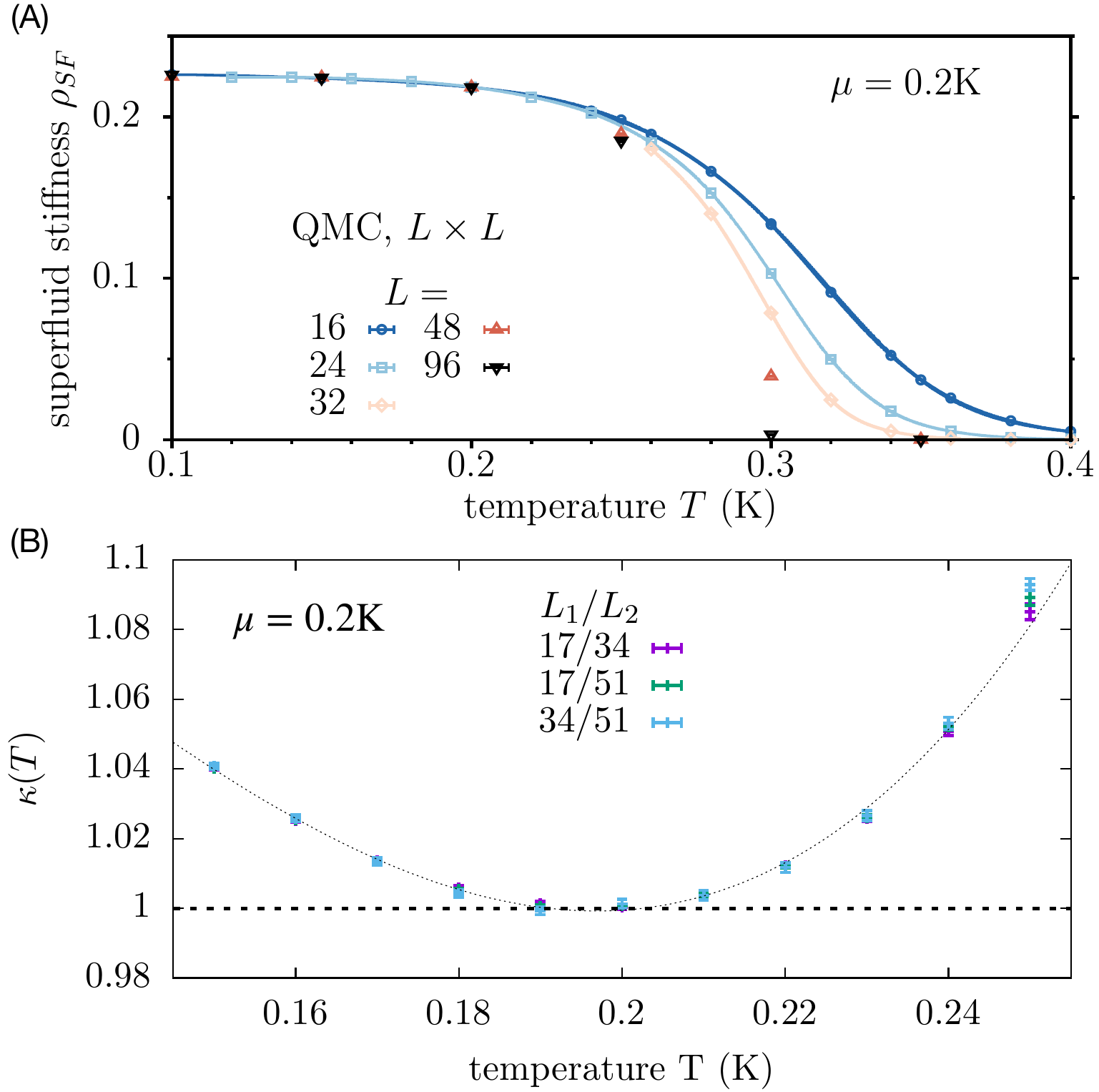}
	\caption{
	(Color online)
	    (A) Superfluid stiffness $\rho_{SF}$ as a function of temperature $T$ at $\mu=0.2$ K for different system sizes.
	     The chemical potential corresponds to magnetic field strengths $H_{m_1}$ of the spin model and is the region with highest condensation temperature $T_c=(0.22\pm0.02)$ K.
	     (B) Microscopic parameter $\kappa(T)$ as a function of temperature $T$ for different system size ratios $L_1/L_2$ for $\mu=0.2$ K. The aspect ratio $L_x/L_y$ of each system size $L_1,L_2$ is chosen such that the superfluid stiffness $\rho_{SF}$ is isotropic.
	}
	\label{fig:BEC_SF}
\end{figure}

In Fig.~\ref{fig:BEC} we show the boson density $n$ as well as the superfluid stiffness $\rho_{SF}$ as a function of chemical potential $\mu$ for a temperature well below the condensation temperature $T_c$.
The density changes from $n=0$ (only singlet states present in the spin model) to $n=1$ for all dimers in their triplet state.
At $T<T_c$ we obtain a finite superfluid stiffness $\rho_{SF}>0$ which indicates a condensation of triplons.
This feature is robust, both for the ferromagnetic (unfrustrated) coupling $J_3=-0.2$ K used in the main paper, and for the antiferromagnetic (frustrated) coupling $J_3=0.2$ K which one would obtain from a fit with $g=2.075$.\\
In order to gain further insights into the condensation temperature $T_c(\mu)$ of the system, the superfluid stiffness needs to be finite-size extrapolated, since $\rho_{SF}$ depends strongly on the system size for $T>T_c$, see Fig.~\ref{fig:BEC_SF}(a).
Following the procedure given in the seminal paper by \textit{Ceperley} and \textit{Pollock} \cite{Ceperley1989} and subsequently used in calculations of the isotropic $t-V$ model on a triangular lattice \cite{Boninsegni2005}, the superfluid stiffness can be extrapolated to the infinite-size limit by using the Kosterlitz-Thouless recursion relations.
In integral form \cite{Prokofev2002}, the renormalization group equation that links two different system sizes $L_1$ and $L_2$ reads 
$$
	\log{L_2/L_1} = \frac{1}{4}\int_{R_2}^{R_1}\frac{\mathrm{d}t}{t^2(\log(t)-\kappa)+t}
$$
with $R=\pi\rho_{SF}/2mT$ defined using the effective mass $m=(t_1+2t_2)^{-1}$.
By using different system size ratios $L_1/L_2$ we determine the microscopic parameter $\kappa$ for different temperatures $T\gtrsim T_c$ in order to determine the temperature $T_c$ at which $\kappa=1$, see Fig.~\ref{fig:BEC_SF}(b).
Error bars for $\kappa$ were calculated based on Monte Carlo error propagation.
It should be noted that the winding number fluctuations $\langle \mathcal{W}^2\rangle$ along the axis of the triangular lattice are usually not the same since we work with an anisotropic model. Since this can have consequences for the precise determination of the critical temperature, we used aspect ratios $\frac{L_x}{L_y}$ which minimize the difference between  $\langle \mathcal{W}_x^2\rangle$ and  $\langle \mathcal{W}_y^2\rangle$ in order to have an isotropic superfluid stiffness $\rho_{SF}$ which allows for a meaningful finite-size scaling as a function of $L$ \cite{You2012}.
The superfluid dome obtained in this way for $J_3=-0.2$ K is shown in the inset of Fig.~4(b) of the main text.
\\
Finally, we note that for $J_3=0.2$ K one obtains $\vert t_2\vert\ll V_2$, which means that the movement of bosons along two directions is frozen out, which for its own would satisfy one of the constraints for forming a supersolid phase.
However, irrespective of the precise value of $J_3$, the fact that dimers couple along the $b$ direction only via $J_1$ always leads to $t_1/V_1=1$ in the effective boson model, hindering the formation of a supersolid phase.
This could be changed by distortions of the perfect in-plane orientation of the dimers along the $a$ direction by out-of-plane buckling and/or in-plane rotations of the dimers.
Thereby, an additional frustrating spin-exchange term between dimers in $a$-direction would occur that would cause a ratio of $t_1/V_1\neq1$ in the boson model.
For sufficiently strong frustration, such a modification of the model could eventually allow for the existence of a supersolid phase.\\
Irrespective of the question on a possible supersolid phase, the precise determination of the spin exchange constant $J_3$ will be important for a detailed discussion of the BEC of triplons and quintuplons in \KNi. 


\begin{thebibliography}{64}%
\makeatletter
\providecommand \@ifxundefined [1]{%
 \@ifx{#1\undefined}
}%
\providecommand \@ifnum [1]{%
 \ifnum #1\expandafter \@firstoftwo
 \else \expandafter \@secondoftwo
 \fi
}%
\providecommand \@ifx [1]{%
 \ifx #1\expandafter \@firstoftwo
 \else \expandafter \@secondoftwo
 \fi
}%
\providecommand \natexlab [1]{#1}%
\providecommand \enquote  [1]{``#1''}%
\providecommand \bibnamefont  [1]{#1}%
\providecommand \bibfnamefont [1]{#1}%
\providecommand \citenamefont [1]{#1}%
\providecommand \href@noop [0]{\@secondoftwo}%
\providecommand \href [0]{\begingroup \@sanitize@url \@href}%
\providecommand \@href[1]{\@@startlink{#1}\@@href}%
\providecommand \@@href[1]{\endgroup#1\@@endlink}%
\providecommand \@sanitize@url [0]{\catcode `\\12\catcode `\$12\catcode
  `\&12\catcode `\#12\catcode `\^12\catcode `\_12\catcode `\%12\relax}%
\providecommand \@@startlink[1]{}%
\providecommand \@@endlink[0]{}%
\providecommand \url  [0]{\begingroup\@sanitize@url \@url }%
\providecommand \@url [1]{\endgroup\@href {#1}{\urlprefix }}%
\providecommand \urlprefix  [0]{URL }%
\providecommand \Eprint [0]{\href }%
\providecommand \doibase [0]{https://doi.org/}%
\providecommand \selectlanguage [0]{\@gobble}%
\providecommand \bibinfo  [0]{\@secondoftwo}%
\providecommand \bibfield  [0]{\@secondoftwo}%
\providecommand \translation [1]{[#1]}%
\providecommand \BibitemOpen [0]{}%
\providecommand \bibitemStop [0]{}%
\providecommand \bibitemNoStop [0]{.\EOS\space}%
\providecommand \EOS [0]{\spacefactor3000\relax}%
\providecommand \BibitemShut  [1]{\csname bibitem#1\endcsname}%
\let\auto@bib@innerbib\@empty
\bibitem [{\citenamefont {Laflorencie}\ and\ \citenamefont
  {Mila}(2007)}]{Laflorencie2007}%
  \BibitemOpen
  \bibfield  {author} {\bibinfo {author} {\bibfnamefont {N.}~\bibnamefont
  {Laflorencie}}\ and\ \bibinfo {author} {\bibfnamefont {F.}~\bibnamefont
  {Mila}},\ }\bibfield  {title} {\bibinfo {title} {{Quantum and Thermal
  Transitions out of the Supersolid Phase of a 2D Quantum Antiferromagnet}},\
  }\href {https://doi.org/10.1103/PhysRevLett.99.027202} {\bibfield  {journal}
  {\bibinfo  {journal} {Phys. Rev. Lett.}\ }\textbf {\bibinfo {volume} {99}},\
  \bibinfo {pages} {027202} (\bibinfo {year} {2007})}\BibitemShut {NoStop}%
\bibitem [{\citenamefont {Sengupta}\ and\ \citenamefont
  {Batista}(2007)}]{Sengupta2007}%
  \BibitemOpen
  \bibfield  {author} {\bibinfo {author} {\bibfnamefont {P.}~\bibnamefont
  {Sengupta}}\ and\ \bibinfo {author} {\bibfnamefont {C.~D.}\ \bibnamefont
  {Batista}},\ }\bibfield  {title} {\bibinfo {title} {{Field-Induced Supersolid
  Phase in Spin-One Heisenberg Models}},\ }\href
  {https://doi.org/10.1103/PhysRevLett.98.227201} {\bibfield  {journal}
  {\bibinfo  {journal} {Phys. Rev. Lett.}\ }\textbf {\bibinfo {volume} {98}},\
  \bibinfo {pages} {227201} (\bibinfo {year} {2007})}\BibitemShut {NoStop}%
\bibitem [{\citenamefont {Greiner}\ \emph {et~al.}(2002)\citenamefont
  {Greiner}, \citenamefont {Mandel}, \citenamefont {Esslinger}, \citenamefont
  {H{\"a}nsch},\ and\ \citenamefont {Bloch}}]{Greiner2002}%
  \BibitemOpen
  \bibfield  {author} {\bibinfo {author} {\bibfnamefont {M.}~\bibnamefont
  {Greiner}}, \bibinfo {author} {\bibfnamefont {O.}~\bibnamefont {Mandel}},
  \bibinfo {author} {\bibfnamefont {T.}~\bibnamefont {Esslinger}}, \bibinfo
  {author} {\bibfnamefont {T.~W.}\ \bibnamefont {H{\"a}nsch}},\ and\ \bibinfo
  {author} {\bibfnamefont {I.}~\bibnamefont {Bloch}},\ }\bibfield  {title}
  {\bibinfo {title} {{Quantum phase transition from a superfluid to a Mott
  insulator in a gas of ultracold atoms}},\ }\href
  {https://doi.org/10.1038/415039a} {\bibfield  {journal} {\bibinfo  {journal}
  {{Nature}}\ }\textbf {\bibinfo {volume} {415}},\ \bibinfo {pages} {39}
  (\bibinfo {year} {2002})}\BibitemShut {NoStop}%
\bibitem [{\citenamefont {Kim}\ and\ \citenamefont
  {Chan}(2004{\natexlab{a}})}]{kim2004a}%
  \BibitemOpen
  \bibfield  {author} {\bibinfo {author} {\bibfnamefont {E.}~\bibnamefont
  {Kim}}\ and\ \bibinfo {author} {\bibfnamefont {M.}~\bibnamefont {Chan}},\
  }\bibfield  {title} {\bibinfo {title} {{Probable observation of a supersolid
  helium phase}},\ }\href {https://doi.org/10.1038/nature02220} {\bibfield
  {journal} {\bibinfo  {journal} {{Nature}}\ }\textbf {\bibinfo {volume}
  {427}},\ \bibinfo {pages} {225} (\bibinfo {year}
  {2004}{\natexlab{a}})}\BibitemShut {NoStop}%
\bibitem [{\citenamefont {Kim}\ and\ \citenamefont
  {Chan}(2004{\natexlab{b}})}]{kim2004b}%
  \BibitemOpen
  \bibfield  {author} {\bibinfo {author} {\bibfnamefont {E.}~\bibnamefont
  {Kim}}\ and\ \bibinfo {author} {\bibfnamefont {M.~H.~W.}\ \bibnamefont
  {Chan}},\ }\bibfield  {title} {\bibinfo {title} {{Observation of Superflow in
  Solid Helium}},\ }\href {https://doi.org/10.1126/science.1101501} {\bibfield
  {journal} {\bibinfo  {journal} {Science}\ }\textbf {\bibinfo {volume}
  {305}},\ \bibinfo {pages} {1941} (\bibinfo {year}
  {2004}{\natexlab{b}})}\BibitemShut {NoStop}%
\bibitem [{\citenamefont {{Giamarchi}}\ \emph {et~al.}(2008)\citenamefont
  {{Giamarchi}}, \citenamefont {{R{\"u}egg}},\ and\ \citenamefont
  {{Tchernyshyov}}}]{Giamarchi2008}%
  \BibitemOpen
  \bibfield  {author} {\bibinfo {author} {\bibfnamefont {T.}~\bibnamefont
  {{Giamarchi}}}, \bibinfo {author} {\bibfnamefont {C.}~\bibnamefont
  {{R{\"u}egg}}},\ and\ \bibinfo {author} {\bibfnamefont {O.}~\bibnamefont
  {{Tchernyshyov}}},\ }\bibfield  {title} {\bibinfo {title} {{Bose-Einstein
  condensation in magnetic insulators}},\ }\href
  {https://doi.org/10.1038/nphys893} {\bibfield  {journal} {\bibinfo  {journal}
  {Nature Physics}\ }\textbf {\bibinfo {volume} {4}},\ \bibinfo {pages} {198}
  (\bibinfo {year} {2008})}\BibitemShut {NoStop}%
\bibitem [{\citenamefont {Zapf}\ \emph {et~al.}(2014)\citenamefont {Zapf},
  \citenamefont {Jaime},\ and\ \citenamefont {Batista}}]{Zapf2014}%
  \BibitemOpen
  \bibfield  {author} {\bibinfo {author} {\bibfnamefont {V.}~\bibnamefont
  {Zapf}}, \bibinfo {author} {\bibfnamefont {M.}~\bibnamefont {Jaime}},\ and\
  \bibinfo {author} {\bibfnamefont {C.~D.}\ \bibnamefont {Batista}},\
  }\bibfield  {title} {\bibinfo {title} {{Bose-Einstein condensation in quantum
  magnets}},\ }\href {https://doi.org/10.1103/RevModPhys.86.563} {\bibfield
  {journal} {\bibinfo  {journal} {Rev. Mod. Phys.}\ }\textbf {\bibinfo {volume}
  {86}},\ \bibinfo {pages} {563} (\bibinfo {year} {2014})}\BibitemShut
  {NoStop}%
\bibitem [{\citenamefont {Matsubara}\ and\ \citenamefont
  {Matsuda}(1956)}]{Matsubara1956}%
  \BibitemOpen
  \bibfield  {author} {\bibinfo {author} {\bibfnamefont {T.}~\bibnamefont
  {Matsubara}}\ and\ \bibinfo {author} {\bibfnamefont {H.}~\bibnamefont
  {Matsuda}},\ }\bibfield  {title} {\bibinfo {title} {{A Lattice Model of
  Liquid Helium, I}},\ }\href {https://doi.org/10.1143/PTP.16.569} {\bibfield
  {journal} {\bibinfo  {journal} {Progress of Theoretical Physics}\ }\textbf
  {\bibinfo {volume} {16}},\ \bibinfo {pages} {569} (\bibinfo {year}
  {1956})}\BibitemShut {NoStop}%
\bibitem [{\citenamefont {Batista}\ and\ \citenamefont
  {Ortiz}(2004)}]{Batista2004}%
  \BibitemOpen
  \bibfield  {author} {\bibinfo {author} {\bibfnamefont {C.~D.}\ \bibnamefont
  {Batista}}\ and\ \bibinfo {author} {\bibfnamefont {G.}~\bibnamefont
  {Ortiz}},\ }\bibfield  {title} {\bibinfo {title} {{Algebraic approach to
  interacting quantum systems}},\ }\href
  {https://doi.org/10.1080/00018730310001642086} {\bibfield  {journal}
  {\bibinfo  {journal} {Advances in Physics}\ }\textbf {\bibinfo {volume}
  {53}},\ \bibinfo {pages} {1} (\bibinfo {year} {2004})}\BibitemShut {NoStop}%
\bibitem [{\citenamefont {Batyev}\ and\ \citenamefont
  {Braginskii}(1984)}]{Batyev84}%
  \BibitemOpen
  \bibfield  {author} {\bibinfo {author} {\bibfnamefont {E.~G.}\ \bibnamefont
  {Batyev}}\ and\ \bibinfo {author} {\bibfnamefont {L.~S.}\ \bibnamefont
  {Braginskii}},\ }\bibfield  {title} {\bibinfo {title} {{Antiferromagnet in a
  strong magnetic field: analogy with Bose gas}},\ }\href@noop {} {\bibfield
  {journal} {\bibinfo  {journal} {Sov. Phys. JETP}\ }\textbf {\bibinfo {volume}
  {60}},\ \bibinfo {pages} {781} (\bibinfo {year} {1984})}\BibitemShut
  {NoStop}%
\bibitem [{\citenamefont {Oosawa}\ \emph {et~al.}(1999)\citenamefont {Oosawa},
  \citenamefont {Ishii},\ and\ \citenamefont {Tanaka}}]{Oosawa1999}%
  \BibitemOpen
  \bibfield  {author} {\bibinfo {author} {\bibfnamefont {A.}~\bibnamefont
  {Oosawa}}, \bibinfo {author} {\bibfnamefont {M.}~\bibnamefont {Ishii}},\ and\
  \bibinfo {author} {\bibfnamefont {H.}~\bibnamefont {Tanaka}},\ }\bibfield
  {title} {\bibinfo {title} {{Field-induced three-dimensional magnetic ordering
  in the spin-gap system}},\ }\href
  {https://doi.org/10.1088/0953-8984/11/1/021} {\bibfield  {journal} {\bibinfo
  {journal} {Journal of Physics: Condensed Matter}\ }\textbf {\bibinfo {volume}
  {11}},\ \bibinfo {pages} {265} (\bibinfo {year} {1999})}\BibitemShut
  {NoStop}%
\bibitem [{\citenamefont {Kageyama}\ \emph {et~al.}(1999)\citenamefont
  {Kageyama}, \citenamefont {Yoshimura}, \citenamefont {Stern}, \citenamefont
  {Mushnikov}, \citenamefont {Onizuka}, \citenamefont {Kato}, \citenamefont
  {Kosuge}, \citenamefont {Slichter}, \citenamefont {Goto},\ and\ \citenamefont
  {Ueda}}]{Kageyama1999}%
  \BibitemOpen
  \bibfield  {author} {\bibinfo {author} {\bibfnamefont {H.}~\bibnamefont
  {Kageyama}}, \bibinfo {author} {\bibfnamefont {K.}~\bibnamefont {Yoshimura}},
  \bibinfo {author} {\bibfnamefont {R.}~\bibnamefont {Stern}}, \bibinfo
  {author} {\bibfnamefont {N.~V.}\ \bibnamefont {Mushnikov}}, \bibinfo {author}
  {\bibfnamefont {K.}~\bibnamefont {Onizuka}}, \bibinfo {author} {\bibfnamefont
  {M.}~\bibnamefont {Kato}}, \bibinfo {author} {\bibfnamefont {K.}~\bibnamefont
  {Kosuge}}, \bibinfo {author} {\bibfnamefont {C.~P.}\ \bibnamefont
  {Slichter}}, \bibinfo {author} {\bibfnamefont {T.}~\bibnamefont {Goto}},\
  and\ \bibinfo {author} {\bibfnamefont {Y.}~\bibnamefont {Ueda}},\ }\bibfield
  {title} {\bibinfo {title} {{Exact Dimer Ground State and Quantized
  Magnetization Plateaus in the Two-Dimensional Spin System
  ${\mathrm{SrCu}}_{2}({\mathrm{BO}}_{3}){}_{2}$}},\ }\href
  {https://doi.org/10.1103/PhysRevLett.82.3168} {\bibfield  {journal} {\bibinfo
   {journal} {Phys. Rev. Lett.}\ }\textbf {\bibinfo {volume} {82}},\ \bibinfo
  {pages} {3168} (\bibinfo {year} {1999})}\BibitemShut {NoStop}%
\bibitem [{\citenamefont {Jaime}\ \emph {et~al.}(2004)\citenamefont {Jaime},
  \citenamefont {Correa}, \citenamefont {Harrison}, \citenamefont {Batista},
  \citenamefont {Kawashima}, \citenamefont {Kazuma}, \citenamefont {Jorge},
  \citenamefont {Stern}, \citenamefont {Heinmaa}, \citenamefont {Zvyagin},
  \citenamefont {Sasago},\ and\ \citenamefont {Uchinokura}}]{Jaime2004}%
  \BibitemOpen
  \bibfield  {author} {\bibinfo {author} {\bibfnamefont {M.}~\bibnamefont
  {Jaime}}, \bibinfo {author} {\bibfnamefont {V.~F.}\ \bibnamefont {Correa}},
  \bibinfo {author} {\bibfnamefont {N.}~\bibnamefont {Harrison}}, \bibinfo
  {author} {\bibfnamefont {C.~D.}\ \bibnamefont {Batista}}, \bibinfo {author}
  {\bibfnamefont {N.}~\bibnamefont {Kawashima}}, \bibinfo {author}
  {\bibfnamefont {Y.}~\bibnamefont {Kazuma}}, \bibinfo {author} {\bibfnamefont
  {G.~A.}\ \bibnamefont {Jorge}}, \bibinfo {author} {\bibfnamefont
  {R.}~\bibnamefont {Stern}}, \bibinfo {author} {\bibfnamefont
  {I.}~\bibnamefont {Heinmaa}}, \bibinfo {author} {\bibfnamefont {S.~A.}\
  \bibnamefont {Zvyagin}}, \bibinfo {author} {\bibfnamefont {Y.}~\bibnamefont
  {Sasago}},\ and\ \bibinfo {author} {\bibfnamefont {K.}~\bibnamefont
  {Uchinokura}},\ }\bibfield  {title} {\bibinfo {title}
  {{Magnetic-Field-Induced Condensation of Triplons in Han Purple Pigment
  ${\mathrm{B}\mathrm{a}\mathrm{C}\mathrm{u}\mathrm{S}\mathrm{i}}_{2}{\mathrm{O}}_{6}$}},\
  }\href {https://doi.org/10.1103/PhysRevLett.93.087203} {\bibfield  {journal}
  {\bibinfo  {journal} {Phys. Rev. Lett.}\ }\textbf {\bibinfo {volume} {93}},\
  \bibinfo {pages} {087203} (\bibinfo {year} {2004})}\BibitemShut {NoStop}%
\bibitem [{\citenamefont {Singh}\ and\ \citenamefont
  {Johnston}(2007)}]{Singh2007}%
  \BibitemOpen
  \bibfield  {author} {\bibinfo {author} {\bibfnamefont {Y.}~\bibnamefont
  {Singh}}\ and\ \bibinfo {author} {\bibfnamefont {D.~C.}\ \bibnamefont
  {Johnston}},\ }\bibfield  {title} {\bibinfo {title} {{Singlet ground state in
  the spin-$\frac{1}{2}$ dimer compound
  ${\mathrm{Sr}}_{3}{\mathrm{Cr}}_{2}{\mathrm{O}}_{8}$}},\ }\href
  {https://doi.org/10.1103/PhysRevB.76.012407} {\bibfield  {journal} {\bibinfo
  {journal} {Phys. Rev. B}\ }\textbf {\bibinfo {volume} {76}},\ \bibinfo
  {pages} {012407} (\bibinfo {year} {2007})}\BibitemShut {NoStop}%
\bibitem [{\citenamefont {Nakajima}\ \emph {et~al.}(2006)\citenamefont
  {Nakajima}, \citenamefont {Mitamura},\ and\ \citenamefont
  {Ueda}}]{Nakajima2006}%
  \BibitemOpen
  \bibfield  {author} {\bibinfo {author} {\bibfnamefont {T.}~\bibnamefont
  {Nakajima}}, \bibinfo {author} {\bibfnamefont {H.}~\bibnamefont {Mitamura}},\
  and\ \bibinfo {author} {\bibfnamefont {Y.}~\bibnamefont {Ueda}},\ }\bibfield
  {title} {\bibinfo {title} {{Singlet Ground State and Magnetic Interactions in
  New Spin Dimer System Ba3Cr2O8}},\ }\href
  {https://doi.org/10.1143/JPSJ.75.054706} {\bibfield  {journal} {\bibinfo
  {journal} {Journal of the Physical Society of Japan}\ }\textbf {\bibinfo
  {volume} {75}},\ \bibinfo {pages} {054706} (\bibinfo {year}
  {2006})}\BibitemShut {NoStop}%
\bibitem [{\citenamefont {Nikuni}\ \emph {et~al.}(2000)\citenamefont {Nikuni},
  \citenamefont {Oshikawa}, \citenamefont {Oosawa},\ and\ \citenamefont
  {Tanaka}}]{Nikuni2000}%
  \BibitemOpen
  \bibfield  {author} {\bibinfo {author} {\bibfnamefont {T.}~\bibnamefont
  {Nikuni}}, \bibinfo {author} {\bibfnamefont {M.}~\bibnamefont {Oshikawa}},
  \bibinfo {author} {\bibfnamefont {A.}~\bibnamefont {Oosawa}},\ and\ \bibinfo
  {author} {\bibfnamefont {H.}~\bibnamefont {Tanaka}},\ }\bibfield  {title}
  {\bibinfo {title} {{Bose-Einstein Condensation of Dilute Magnons in
  ${\mathrm{TlCuCl}}_{3}$}},\ }\href
  {https://doi.org/10.1103/PhysRevLett.84.5868} {\bibfield  {journal} {\bibinfo
   {journal} {Phys. Rev. Lett.}\ }\textbf {\bibinfo {volume} {84}},\ \bibinfo
  {pages} {5868} (\bibinfo {year} {2000})}\BibitemShut {NoStop}%
\bibitem [{\citenamefont {R\"uegg}\ \emph {et~al.}(2003)\citenamefont
  {R\"uegg}, \citenamefont {Cavadini}, \citenamefont {Furrer}, \citenamefont
  {G\"udel}, \citenamefont {Kr\"amer}, \citenamefont {Mutka}, \citenamefont
  {Wildes}, \citenamefont {Habicht},\ and\ \citenamefont
  {Vorderwisch}}]{Ruegg2003}%
  \BibitemOpen
  \bibfield  {author} {\bibinfo {author} {\bibfnamefont {C.}~\bibnamefont
  {R\"uegg}}, \bibinfo {author} {\bibfnamefont {N.}~\bibnamefont {Cavadini}},
  \bibinfo {author} {\bibfnamefont {A.}~\bibnamefont {Furrer}}, \bibinfo
  {author} {\bibfnamefont {H.-J.}\ \bibnamefont {G\"udel}}, \bibinfo {author}
  {\bibfnamefont {K.}~\bibnamefont {Kr\"amer}}, \bibinfo {author}
  {\bibfnamefont {H.}~\bibnamefont {Mutka}}, \bibinfo {author} {\bibfnamefont
  {A.}~\bibnamefont {Wildes}}, \bibinfo {author} {\bibfnamefont
  {K.}~\bibnamefont {Habicht}},\ and\ \bibinfo {author} {\bibfnamefont
  {P.}~\bibnamefont {Vorderwisch}},\ }\bibfield  {title} {\bibinfo {title}
  {{Bose-Einstein condensation of the triplet states in the magnetic insulator
  TlCuCl$_3$}},\ }\href {http://dx.doi.org/10.1038/nature01617} {\bibfield
  {journal} {\bibinfo  {journal} {{Nature}}\ }\textbf {\bibinfo {volume}
  {423}},\ \bibinfo {pages} {62} (\bibinfo {year} {2003})}\BibitemShut
  {NoStop}%
\bibitem [{\citenamefont {Aczel}\ \emph {et~al.}(2009)\citenamefont {Aczel},
  \citenamefont {Kohama}, \citenamefont {Jaime}, \citenamefont {Ninios},
  \citenamefont {Chan}, \citenamefont {Balicas}, \citenamefont {Dabkowska},\
  and\ \citenamefont {Luke}}]{Aczel2009}%
  \BibitemOpen
  \bibfield  {author} {\bibinfo {author} {\bibfnamefont {A.~A.}\ \bibnamefont
  {Aczel}}, \bibinfo {author} {\bibfnamefont {Y.}~\bibnamefont {Kohama}},
  \bibinfo {author} {\bibfnamefont {M.}~\bibnamefont {Jaime}}, \bibinfo
  {author} {\bibfnamefont {K.}~\bibnamefont {Ninios}}, \bibinfo {author}
  {\bibfnamefont {H.~B.}\ \bibnamefont {Chan}}, \bibinfo {author}
  {\bibfnamefont {L.}~\bibnamefont {Balicas}}, \bibinfo {author} {\bibfnamefont
  {H.~A.}\ \bibnamefont {Dabkowska}},\ and\ \bibinfo {author} {\bibfnamefont
  {G.~M.}\ \bibnamefont {Luke}},\ }\bibfield  {title} {\bibinfo {title}
  {{Bose-Einstein condensation of triplons in
  ${\text{Ba}}_{3}{\text{Cr}}_{2}{\text{O}}_{8}$}},\ }\href
  {https://doi.org/10.1103/PhysRevB.79.100409} {\bibfield  {journal} {\bibinfo
  {journal} {Phys. Rev. B}\ }\textbf {\bibinfo {volume} {79}},\ \bibinfo
  {pages} {100409} (\bibinfo {year} {2009})}\BibitemShut {NoStop}%
\bibitem [{\citenamefont {Uchida}\ \emph {et~al.}(2001)\citenamefont {Uchida},
  \citenamefont {Tanaka}, \citenamefont {Bartashevich},\ and\ \citenamefont
  {Goto}}]{Uchida2001}%
  \BibitemOpen
  \bibfield  {author} {\bibinfo {author} {\bibfnamefont {M.}~\bibnamefont
  {Uchida}}, \bibinfo {author} {\bibfnamefont {H.}~\bibnamefont {Tanaka}},
  \bibinfo {author} {\bibfnamefont {M.}~\bibnamefont {Bartashevich}},\ and\
  \bibinfo {author} {\bibfnamefont {T.}~\bibnamefont {Goto}},\ }\bibfield
  {title} {\bibinfo {title} {{Singlet Ground State and Magnetization Plateaus
  in Ba$_3$Mn$_2$O$_8$}},\ }\href {https://doi.org/10.1143/JPSJ.70.1790}
  {\bibfield  {journal} {\bibinfo  {journal} {Journal of the Physical Society
  of Japan}\ }\textbf {\bibinfo {volume} {70}},\ \bibinfo {pages} {1790}
  (\bibinfo {year} {2001})}\BibitemShut {NoStop}%
\bibitem [{\citenamefont {Hosokoshi}\ \emph {et~al.}(1999)\citenamefont
  {Hosokoshi}, \citenamefont {Nakazawa}, \citenamefont {Inoue}, \citenamefont
  {Takizawa}, \citenamefont {Nakano}, \citenamefont {Takahashi},\ and\
  \citenamefont {Goto}}]{Hosokoshi1999}%
  \BibitemOpen
  \bibfield  {author} {\bibinfo {author} {\bibfnamefont {Y.}~\bibnamefont
  {Hosokoshi}}, \bibinfo {author} {\bibfnamefont {Y.}~\bibnamefont {Nakazawa}},
  \bibinfo {author} {\bibfnamefont {K.}~\bibnamefont {Inoue}}, \bibinfo
  {author} {\bibfnamefont {K.}~\bibnamefont {Takizawa}}, \bibinfo {author}
  {\bibfnamefont {H.}~\bibnamefont {Nakano}}, \bibinfo {author} {\bibfnamefont
  {M.}~\bibnamefont {Takahashi}},\ and\ \bibinfo {author} {\bibfnamefont
  {T.}~\bibnamefont {Goto}},\ }\bibfield  {title} {\bibinfo {title} {{Magnetic
  properties of low-dimensional quantum spin systems made of stable organic
  biradicals PNNNO, ${\mathrm{F}}_{2}\mathrm{PNNNO},$ and PIMNO}},\ }\href
  {https://doi.org/10.1103/PhysRevB.60.12924} {\bibfield  {journal} {\bibinfo
  {journal} {Phys. Rev. B}\ }\textbf {\bibinfo {volume} {60}},\ \bibinfo
  {pages} {12924} (\bibinfo {year} {1999})}\BibitemShut {NoStop}%
\bibitem [{\citenamefont {Uchida}\ \emph {et~al.}(2002)\citenamefont {Uchida},
  \citenamefont {Tanaka}, \citenamefont {Mitamura}, \citenamefont {Ishikawa},\
  and\ \citenamefont {Goto}}]{Uchida2002}%
  \BibitemOpen
  \bibfield  {author} {\bibinfo {author} {\bibfnamefont {M.}~\bibnamefont
  {Uchida}}, \bibinfo {author} {\bibfnamefont {H.}~\bibnamefont {Tanaka}},
  \bibinfo {author} {\bibfnamefont {H.}~\bibnamefont {Mitamura}}, \bibinfo
  {author} {\bibfnamefont {F.}~\bibnamefont {Ishikawa}},\ and\ \bibinfo
  {author} {\bibfnamefont {T.}~\bibnamefont {Goto}},\ }\bibfield  {title}
  {\bibinfo {title} {{High-field magnetization process in the $S=1$ quantum
  spin system ${\mathrm{Ba}}_{3}{\mathrm{Mn}}_{2}{\mathrm{O}}_{8}$}},\ }\href
  {https://doi.org/10.1103/PhysRevB.66.054429} {\bibfield  {journal} {\bibinfo
  {journal} {Phys. Rev. B}\ }\textbf {\bibinfo {volume} {66}},\ \bibinfo
  {pages} {054429} (\bibinfo {year} {2002})}\BibitemShut {NoStop}%
\bibitem [{\citenamefont {Tsujii}\ \emph {et~al.}(2005)\citenamefont {Tsujii},
  \citenamefont {Andraka}, \citenamefont {Uchida}, \citenamefont {Tanaka},\
  and\ \citenamefont {Takano}}]{Tsuji2005}%
  \BibitemOpen
  \bibfield  {author} {\bibinfo {author} {\bibfnamefont {H.}~\bibnamefont
  {Tsujii}}, \bibinfo {author} {\bibfnamefont {B.}~\bibnamefont {Andraka}},
  \bibinfo {author} {\bibfnamefont {M.}~\bibnamefont {Uchida}}, \bibinfo
  {author} {\bibfnamefont {H.}~\bibnamefont {Tanaka}},\ and\ \bibinfo {author}
  {\bibfnamefont {Y.}~\bibnamefont {Takano}},\ }\bibfield  {title} {\bibinfo
  {title} {{Specific heat of the $S=1$ spin-dimer antiferromagnet
  ${\mathrm{Ba}}_{3}{\mathrm{Mn}}_{2}{\mathrm{O}}_{8}$ in high magnetic
  fields}},\ }\href {https://doi.org/10.1103/PhysRevB.72.214434} {\bibfield
  {journal} {\bibinfo  {journal} {Phys. Rev. B}\ }\textbf {\bibinfo {volume}
  {72}},\ \bibinfo {pages} {214434} (\bibinfo {year} {2005})}\BibitemShut
  {NoStop}%
\bibitem [{\citenamefont {Stone}\ \emph {et~al.}(2008)\citenamefont {Stone},
  \citenamefont {Lumsden}, \citenamefont {Chang}, \citenamefont {Samulon},
  \citenamefont {Batista},\ and\ \citenamefont {Fisher}}]{Stone2008}%
  \BibitemOpen
  \bibfield  {author} {\bibinfo {author} {\bibfnamefont {M.~B.}\ \bibnamefont
  {Stone}}, \bibinfo {author} {\bibfnamefont {M.~D.}\ \bibnamefont {Lumsden}},
  \bibinfo {author} {\bibfnamefont {S.}~\bibnamefont {Chang}}, \bibinfo
  {author} {\bibfnamefont {E.~C.}\ \bibnamefont {Samulon}}, \bibinfo {author}
  {\bibfnamefont {C.~D.}\ \bibnamefont {Batista}},\ and\ \bibinfo {author}
  {\bibfnamefont {I.~R.}\ \bibnamefont {Fisher}},\ }\bibfield  {title}
  {\bibinfo {title} {{Singlet-Triplet Dispersion Reveals Additional Frustration
  in the Triangular-Lattice Dimer Compound
  ${\mathrm{Ba}}_{3}{\mathrm{Mn}}_{2}{\mathrm{O}}_{8}$}},\ }\href
  {https://doi.org/10.1103/PhysRevLett.100.237201} {\bibfield  {journal}
  {\bibinfo  {journal} {Phys. Rev. Lett.}\ }\textbf {\bibinfo {volume} {100}},\
  \bibinfo {pages} {237201} (\bibinfo {year} {2008})}\BibitemShut {NoStop}%
\bibitem [{\citenamefont {Samulon}\ \emph {et~al.}(2008)\citenamefont
  {Samulon}, \citenamefont {Jo}, \citenamefont {Sengupta}, \citenamefont
  {Batista}, \citenamefont {Jaime}, \citenamefont {Balicas},\ and\
  \citenamefont {Fisher}}]{Samulon2008}%
  \BibitemOpen
  \bibfield  {author} {\bibinfo {author} {\bibfnamefont {E.~C.}\ \bibnamefont
  {Samulon}}, \bibinfo {author} {\bibfnamefont {Y.-J.}\ \bibnamefont {Jo}},
  \bibinfo {author} {\bibfnamefont {P.}~\bibnamefont {Sengupta}}, \bibinfo
  {author} {\bibfnamefont {C.~D.}\ \bibnamefont {Batista}}, \bibinfo {author}
  {\bibfnamefont {M.}~\bibnamefont {Jaime}}, \bibinfo {author} {\bibfnamefont
  {L.}~\bibnamefont {Balicas}},\ and\ \bibinfo {author} {\bibfnamefont {I.~R.}\
  \bibnamefont {Fisher}},\ }\bibfield  {title} {\bibinfo {title} {{Ordered
  magnetic phases of the frustrated spin-dimer compound
  ${\text{Ba}}_{3}{\text{Mn}}_{2}{\text{O}}_{8}$}},\ }\href
  {https://doi.org/10.1103/PhysRevB.77.214441} {\bibfield  {journal} {\bibinfo
  {journal} {Phys. Rev. B}\ }\textbf {\bibinfo {volume} {77}},\ \bibinfo
  {pages} {214441} (\bibinfo {year} {2008})}\BibitemShut {NoStop}%
\bibitem [{\citenamefont {Samulon}\ \emph {et~al.}(2009)\citenamefont
  {Samulon}, \citenamefont {Kohama}, \citenamefont {McDonald}, \citenamefont
  {Shapiro}, \citenamefont {Al-Hassanieh}, \citenamefont {Batista},
  \citenamefont {Jaime},\ and\ \citenamefont {Fisher}}]{Samulon2009}%
  \BibitemOpen
  \bibfield  {author} {\bibinfo {author} {\bibfnamefont {E.~C.}\ \bibnamefont
  {Samulon}}, \bibinfo {author} {\bibfnamefont {Y.}~\bibnamefont {Kohama}},
  \bibinfo {author} {\bibfnamefont {R.~D.}\ \bibnamefont {McDonald}}, \bibinfo
  {author} {\bibfnamefont {M.~C.}\ \bibnamefont {Shapiro}}, \bibinfo {author}
  {\bibfnamefont {K.~A.}\ \bibnamefont {Al-Hassanieh}}, \bibinfo {author}
  {\bibfnamefont {C.~D.}\ \bibnamefont {Batista}}, \bibinfo {author}
  {\bibfnamefont {M.}~\bibnamefont {Jaime}},\ and\ \bibinfo {author}
  {\bibfnamefont {I.~R.}\ \bibnamefont {Fisher}},\ }\bibfield  {title}
  {\bibinfo {title} {{Asymmetric Quintuplet Condensation in the Frustrated
  $S=1$ Spin Dimer Compound
  ${\mathrm{Ba}}_{3}{\mathrm{Mn}}_{2}{\mathbf{O}}_{8}$}},\ }\href
  {https://doi.org/10.1103/PhysRevLett.103.047202} {\bibfield  {journal}
  {\bibinfo  {journal} {Phys. Rev. Lett.}\ }\textbf {\bibinfo {volume} {103}},\
  \bibinfo {pages} {047202} (\bibinfo {year} {2009})}\BibitemShut {NoStop}%
\bibitem [{\citenamefont {Samulon}\ \emph {et~al.}(2010)\citenamefont
  {Samulon}, \citenamefont {Al-Hassanieh}, \citenamefont {Jo}, \citenamefont
  {Shapiro}, \citenamefont {Balicas}, \citenamefont {Batista},\ and\
  \citenamefont {Fisher}}]{Samulon2010}%
  \BibitemOpen
  \bibfield  {author} {\bibinfo {author} {\bibfnamefont {E.~C.}\ \bibnamefont
  {Samulon}}, \bibinfo {author} {\bibfnamefont {K.~A.}\ \bibnamefont
  {Al-Hassanieh}}, \bibinfo {author} {\bibfnamefont {Y.-J.}\ \bibnamefont
  {Jo}}, \bibinfo {author} {\bibfnamefont {M.~C.}\ \bibnamefont {Shapiro}},
  \bibinfo {author} {\bibfnamefont {L.}~\bibnamefont {Balicas}}, \bibinfo
  {author} {\bibfnamefont {C.~D.}\ \bibnamefont {Batista}},\ and\ \bibinfo
  {author} {\bibfnamefont {I.~R.}\ \bibnamefont {Fisher}},\ }\bibfield  {title}
  {\bibinfo {title} {{Anisotropic phase diagram of the frustrated spin dimer
  compound ${\text{Ba}}_{3}{\text{Mn}}_{2}{\text{O}}_{8}$}},\ }\href
  {https://doi.org/10.1103/PhysRevB.81.104421} {\bibfield  {journal} {\bibinfo
  {journal} {Phys. Rev. B}\ }\textbf {\bibinfo {volume} {81}},\ \bibinfo
  {pages} {104421} (\bibinfo {year} {2010})}\BibitemShut {NoStop}%
\bibitem [{\citenamefont {Murugan}\ \emph {et~al.}(2021)\citenamefont
  {Murugan}, \citenamefont {Babu}, \citenamefont {Sankar}, \citenamefont
  {Chen}, \citenamefont {Muthuselvam}, \citenamefont {Chattopadhyay},\ and\
  \citenamefont {Choi}}]{Murugan2021}%
  \BibitemOpen
  \bibfield  {author} {\bibinfo {author} {\bibfnamefont {G.~S.}\ \bibnamefont
  {Murugan}}, \bibinfo {author} {\bibfnamefont {K.~R.}\ \bibnamefont {Babu}},
  \bibinfo {author} {\bibfnamefont {R.}~\bibnamefont {Sankar}}, \bibinfo
  {author} {\bibfnamefont {W.~T.}\ \bibnamefont {Chen}}, \bibinfo {author}
  {\bibfnamefont {I.~P.}\ \bibnamefont {Muthuselvam}}, \bibinfo {author}
  {\bibfnamefont {S.}~\bibnamefont {Chattopadhyay}},\ and\ \bibinfo {author}
  {\bibfnamefont {K.-Y.}\ \bibnamefont {Choi}},\ }\bibfield  {title} {\bibinfo
  {title} {{Magnetic and structural dimer networks in layered
  ${\mathrm{K}}_{2}\mathrm{Ni}$(${\mathrm{MoO}}_{4}{)}_{2}$}},\ }\href
  {https://doi.org/10.1103/PhysRevB.103.024451} {\bibfield  {journal} {\bibinfo
   {journal} {Phys. Rev. B}\ }\textbf {\bibinfo {volume} {103}},\ \bibinfo
  {pages} {024451} (\bibinfo {year} {2021})}\BibitemShut {NoStop}%
\bibitem [{\citenamefont {Klevtsova}\ and\ \citenamefont
  {Klevtsov}(1978)}]{Klevtsova1978}%
  \BibitemOpen
  \bibfield  {author} {\bibinfo {author} {\bibfnamefont {R.}~\bibnamefont
  {Klevtsova}}\ and\ \bibinfo {author} {\bibfnamefont {P.}~\bibnamefont
  {Klevtsov}},\ }\bibfield  {title} {\bibinfo {title} {{Crystal structure of
  double molybdate K$_2$Ni(MoO$_4$)$_2$}},\ }\href
  {http://inis.iaea.org/search/search.aspx?orig_q=RN:10444040} {\bibfield
  {journal} {\bibinfo  {journal} {Kristallografiya}\ }\textbf {\bibinfo
  {volume} {23}},\ \bibinfo {pages} {261} (\bibinfo {year} {1978})}\BibitemShut
  {NoStop}%
\bibitem [{\citenamefont {Hohenberg}\ and\ \citenamefont {Kohn}(1964)}]{DFT1}%
  \BibitemOpen
  \bibfield  {author} {\bibinfo {author} {\bibfnamefont {P.}~\bibnamefont
  {Hohenberg}}\ and\ \bibinfo {author} {\bibfnamefont {W.}~\bibnamefont
  {Kohn}},\ }\bibfield  {title} {\bibinfo {title} {{Inhomogeneous Electron
  Gas}},\ }\href {https://doi.org/10.1103/PhysRev.136.B864} {\bibfield
  {journal} {\bibinfo  {journal} {Phys. Rev.}\ }\textbf {\bibinfo {volume}
  {136}},\ \bibinfo {pages} {B864} (\bibinfo {year} {1964})}\BibitemShut
  {NoStop}%
\bibitem [{\citenamefont {Jones}\ and\ \citenamefont
  {Gunnarsson}(1989)}]{DFT2}%
  \BibitemOpen
  \bibfield  {author} {\bibinfo {author} {\bibfnamefont {R.~O.}\ \bibnamefont
  {Jones}}\ and\ \bibinfo {author} {\bibfnamefont {O.}~\bibnamefont
  {Gunnarsson}},\ }\bibfield  {title} {\bibinfo {title} {{The density
  functional formalism, its applications and prospects}},\ }\href
  {https://doi.org/10.1103/RevModPhys.61.689} {\bibfield  {journal} {\bibinfo
  {journal} {Rev. Mod. Phys.}\ }\textbf {\bibinfo {volume} {61}},\ \bibinfo
  {pages} {689} (\bibinfo {year} {1989})}\BibitemShut {NoStop}%
\bibitem [{\citenamefont {Marzari}\ \emph {et~al.}(2012)\citenamefont
  {Marzari}, \citenamefont {Mostofi}, \citenamefont {Yates}, \citenamefont
  {Souza},\ and\ \citenamefont {Vanderbilt}}]{Wannier}%
  \BibitemOpen
  \bibfield  {author} {\bibinfo {author} {\bibfnamefont {N.}~\bibnamefont
  {Marzari}}, \bibinfo {author} {\bibfnamefont {A.~A.}\ \bibnamefont
  {Mostofi}}, \bibinfo {author} {\bibfnamefont {J.~R.}\ \bibnamefont {Yates}},
  \bibinfo {author} {\bibfnamefont {I.}~\bibnamefont {Souza}},\ and\ \bibinfo
  {author} {\bibfnamefont {D.}~\bibnamefont {Vanderbilt}},\ }\bibfield  {title}
  {\bibinfo {title} {{Maximally localized Wannier functions: Theory and
  applications}},\ }\href {https://doi.org/10.1103/RevModPhys.84.1419}
  {\bibfield  {journal} {\bibinfo  {journal} {Rev. Mod. Phys.}\ }\textbf
  {\bibinfo {volume} {84}},\ \bibinfo {pages} {1419} (\bibinfo {year}
  {2012})}\BibitemShut {NoStop}%
\bibitem [{Sup()}]{SupMat}%
  \BibitemOpen
  \href@noop {} {\bibinfo  {journal} {{See Supplemental Material, which also
  includes Refs.\cite{,Loewdin1951,FPLMTOCode,Exchange_Rspt,Liechtenstein1995,Blochl1994,Kresse1996,Blaha1999,FPLMTO,FPLMTO_Orig,PhysRevB.96.045137,fullprof,wannier90,Wien2wannier,Boninsegni2005,
  Wessel2005, Heidarian2005, Melko2005, Gan2008, Zhang2016,Zvyagin2006}}}\
  }\BibitemShut {NoStop}%
\bibitem [{Note1()}]{Note1}%
  \BibitemOpen
  \bibinfo {note} {Details of the fit and our mean-field estimate of the
  inter-dimer coupling, which is slightly different from Ref.~\protect
  \rev@citealp {Murugan2021}, can be found in the supplemental information
  \cite {SupMat}}\BibitemShut {NoStop}%
\bibitem [{\citenamefont {Anisimov}\ \emph {et~al.}(1997)\citenamefont
  {Anisimov}, \citenamefont {Aryasetiawan},\ and\ \citenamefont
  {Lichtenstein}}]{LSDAU_FLL}%
  \BibitemOpen
\bibfield  {journal} {  }\bibfield  {author} {\bibinfo {author} {\bibfnamefont
  {V.~I.}\ \bibnamefont {Anisimov}}, \bibinfo {author} {\bibfnamefont
  {F.}~\bibnamefont {Aryasetiawan}},\ and\ \bibinfo {author} {\bibfnamefont
  {A.~I.}\ \bibnamefont {Lichtenstein}},\ }\bibfield  {title} {\bibinfo {title}
  {{First-principles calculations of the electronic structure and spectra of
  strongly correlated systems: the LDA+U method}},\ }\href
  {https://doi.org/10.1088/0953-8984/9/4/002} {\bibfield  {journal} {\bibinfo
  {journal} {J. Phys.: Condens. Matter}\ }\textbf {\bibinfo {volume} {9}},\
  \bibinfo {pages} {767} (\bibinfo {year} {1997})}\BibitemShut {NoStop}%
\bibitem [{\citenamefont {Liechtenstein}\ \emph {et~al.}(1987)\citenamefont
  {Liechtenstein}, \citenamefont {Katsnelson}, \citenamefont {Antropov},\ and\
  \citenamefont {Gubanov}}]{magneticforceth1}%
  \BibitemOpen
  \bibfield  {author} {\bibinfo {author} {\bibfnamefont {A.}~\bibnamefont
  {Liechtenstein}}, \bibinfo {author} {\bibfnamefont {M.}~\bibnamefont
  {Katsnelson}}, \bibinfo {author} {\bibfnamefont {V.}~\bibnamefont
  {Antropov}},\ and\ \bibinfo {author} {\bibfnamefont {V.}~\bibnamefont
  {Gubanov}},\ }\bibfield  {title} {\bibinfo {title} {{Local spin density
  functional approach to the theory of exchange interactions in ferromagnetic
  metals and alloys}},\ }\href {https://doi.org/10.1016/0304-8853(87)90721-9}
  {\bibfield  {journal} {\bibinfo  {journal} {Journal of Magnetism and Magnetic
  Materials}\ }\textbf {\bibinfo {volume} {67}},\ \bibinfo {pages} {65 }
  (\bibinfo {year} {1987})}\BibitemShut {NoStop}%
\bibitem [{\citenamefont {Katsnelson}\ and\ \citenamefont
  {Lichtenstein}(2000)}]{magneticforceth2}%
  \BibitemOpen
  \bibfield  {author} {\bibinfo {author} {\bibfnamefont {M.~I.}\ \bibnamefont
  {Katsnelson}}\ and\ \bibinfo {author} {\bibfnamefont {A.~I.}\ \bibnamefont
  {Lichtenstein}},\ }\bibfield  {title} {\bibinfo {title} {{First-principles
  calculations of magnetic interactions in correlated systems}},\ }\href
  {https://doi.org/10.1103/PhysRevB.61.8906} {\bibfield  {journal} {\bibinfo
  {journal} {Phys. Rev. B}\ }\textbf {\bibinfo {volume} {61}},\ \bibinfo
  {pages} {8906} (\bibinfo {year} {2000})}\BibitemShut {NoStop}%
\bibitem [{\citenamefont {Prokof'ev}\ \emph {et~al.}(1998)\citenamefont
  {Prokof'ev}, \citenamefont {Svistunov},\ and\ \citenamefont
  {Tupitsyn}}]{Prokofev1998}%
  \BibitemOpen
  \bibfield  {author} {\bibinfo {author} {\bibfnamefont {N.}~\bibnamefont
  {Prokof'ev}}, \bibinfo {author} {\bibfnamefont {B.}~\bibnamefont
  {Svistunov}},\ and\ \bibinfo {author} {\bibfnamefont {I.}~\bibnamefont
  {Tupitsyn}},\ }\bibfield  {title} {\bibinfo {title} {{Worm algorithm in
  quantum Monte Carlo simulations}},\ }\href
  {https://doi.org/https://doi.org/10.1016/S0375-9601(97)00957-2} {\bibfield
  {journal} {\bibinfo  {journal} {Physics Letters A}\ }\textbf {\bibinfo
  {volume} {238}},\ \bibinfo {pages} {253 } (\bibinfo {year}
  {1998})}\BibitemShut {NoStop}%
\bibitem [{\citenamefont {{Prokof'ev, N. V. and Svistunov, B. V. and Tupitsyn,
  I. S.}}(1998)}]{Prokofev1998b}%
  \BibitemOpen
  \bibfield  {author} {\bibinfo {author} {\bibnamefont {{Prokof'ev, N. V. and
  Svistunov, B. V. and Tupitsyn, I. S.}}},\ }\bibfield  {title} {\bibinfo
  {title} {{Exact, complete, and universal continuous-time worldline Monte
  Carlo approach to the statistics of discrete quantum systems}},\ }\href
  {https://doi.org/10.1134/1.558661} {\bibfield  {journal} {\bibinfo  {journal}
  {Journal of Experimental and Theoretical Physics}\ }\textbf {\bibinfo
  {volume} {87}},\ \bibinfo {pages} {310} (\bibinfo {year} {1998})}\BibitemShut
  {NoStop}%
\bibitem [{\citenamefont {{Troyer, M. and Alet, F. and Trebst, S. and Wessel,
  S.}}(2003)}]{Troyer2003}%
  \BibitemOpen
  \bibfield  {author} {\bibinfo {author} {\bibnamefont {{Troyer, M. and Alet,
  F. and Trebst, S. and Wessel, S.}}},\ }\bibfield  {title} {\bibinfo {title}
  {{Non-local Updates for Quantum Monte Carlo Simulations}},\ }\href
  {https://doi.org/10.1063/1.1632126} {\bibfield  {journal} {\bibinfo
  {journal} {AIP Conference Proceedings}\ }\textbf {\bibinfo {volume} {690}},\
  \bibinfo {pages} {156} (\bibinfo {year} {2003})}\BibitemShut {NoStop}%
\bibitem [{\citenamefont {Albuquerque}\ \emph {et~al.}(2007)\citenamefont
  {Albuquerque}, \citenamefont {Alet}, \citenamefont {Corboz}, \citenamefont
  {Dayal}, \citenamefont {Feiguin}, \citenamefont {Fuchs}, \citenamefont
  {Gamper}, \citenamefont {Gull}, \citenamefont {G\"urtler}, \citenamefont
  {Honecker}, \citenamefont {Igarashi}, \citenamefont {Körner}, \citenamefont
  {Kozhevnikov}, \citenamefont {L\"auchli}, \citenamefont {Manmana},
  \citenamefont {Matsumoto}, \citenamefont {McCulloch}, \citenamefont {Michel},
  \citenamefont {Noack}, \citenamefont {Pawlowski}, \citenamefont {Pollet},
  \citenamefont {Pruschke}, \citenamefont {Schollw\"ock}, \citenamefont {Todo},
  \citenamefont {Trebst}, \citenamefont {Troyer}, \citenamefont {Werner},\ and\
  \citenamefont {Wessel}}]{ALPS1}%
  \BibitemOpen
  \bibfield  {author} {\bibinfo {author} {\bibfnamefont {A.}~\bibnamefont
  {Albuquerque}}, \bibinfo {author} {\bibfnamefont {F.}~\bibnamefont {Alet}},
  \bibinfo {author} {\bibfnamefont {P.}~\bibnamefont {Corboz}}, \bibinfo
  {author} {\bibfnamefont {P.}~\bibnamefont {Dayal}}, \bibinfo {author}
  {\bibfnamefont {A.}~\bibnamefont {Feiguin}}, \bibinfo {author} {\bibfnamefont
  {S.}~\bibnamefont {Fuchs}}, \bibinfo {author} {\bibfnamefont
  {L.}~\bibnamefont {Gamper}}, \bibinfo {author} {\bibfnamefont
  {E.}~\bibnamefont {Gull}}, \bibinfo {author} {\bibfnamefont {S.}~\bibnamefont
  {G\"urtler}}, \bibinfo {author} {\bibfnamefont {A.}~\bibnamefont {Honecker}},
  \bibinfo {author} {\bibfnamefont {R.}~\bibnamefont {Igarashi}}, \bibinfo
  {author} {\bibfnamefont {M.}~\bibnamefont {Körner}}, \bibinfo {author}
  {\bibfnamefont {A.}~\bibnamefont {Kozhevnikov}}, \bibinfo {author}
  {\bibfnamefont {A.}~\bibnamefont {L\"auchli}}, \bibinfo {author}
  {\bibfnamefont {S.}~\bibnamefont {Manmana}}, \bibinfo {author} {\bibfnamefont
  {M.}~\bibnamefont {Matsumoto}}, \bibinfo {author} {\bibfnamefont
  {I.}~\bibnamefont {McCulloch}}, \bibinfo {author} {\bibfnamefont
  {F.}~\bibnamefont {Michel}}, \bibinfo {author} {\bibfnamefont
  {R.}~\bibnamefont {Noack}}, \bibinfo {author} {\bibfnamefont
  {G.}~\bibnamefont {Pawlowski}}, \bibinfo {author} {\bibfnamefont
  {L.}~\bibnamefont {Pollet}}, \bibinfo {author} {\bibfnamefont
  {T.}~\bibnamefont {Pruschke}}, \bibinfo {author} {\bibfnamefont
  {U.}~\bibnamefont {Schollw\"ock}}, \bibinfo {author} {\bibfnamefont
  {S.}~\bibnamefont {Todo}}, \bibinfo {author} {\bibfnamefont {S.}~\bibnamefont
  {Trebst}}, \bibinfo {author} {\bibfnamefont {M.}~\bibnamefont {Troyer}},
  \bibinfo {author} {\bibfnamefont {P.}~\bibnamefont {Werner}},\ and\ \bibinfo
  {author} {\bibfnamefont {S.}~\bibnamefont {Wessel}},\ }\bibfield  {title}
  {\bibinfo {title} {{The ALPS project release 1.3: Open-source software for
  strongly correlated systems}},\ }\href
  {https://doi.org/https://doi.org/10.1016/j.jmmm.2006.10.304} {\bibfield
  {journal} {\bibinfo  {journal} {{Journal of Magnetism and Magnetic
  Materials}}\ }\textbf {\bibinfo {volume} {310}},\ \bibinfo {pages} {1187 }
  (\bibinfo {year} {2007})},\ \bibinfo {note} {{Proceedings of the 17th
  International Conference on Magnetism}}\BibitemShut {NoStop}%
\bibitem [{\citenamefont {Bauer}\ \emph {et~al.}(2011)\citenamefont {Bauer},
  \citenamefont {Carr}, \citenamefont {Evertz}, \citenamefont {Feiguin},
  \citenamefont {Freire}, \citenamefont {Fuchs}, \citenamefont {Gamper},
  \citenamefont {Gukelberger}, \citenamefont {Gull}, \citenamefont {Guertler},
  \citenamefont {Hehn}, \citenamefont {Igarashi}, \citenamefont {Isakov},
  \citenamefont {Koop}, \citenamefont {Ma}, \citenamefont {Mates},
  \citenamefont {Matsuo}, \citenamefont {Parcollet}, \citenamefont {Pawlowski},
  \citenamefont {Picon}, \citenamefont {Pollet}, \citenamefont {Santos},
  \citenamefont {Scarola}, \citenamefont {Schollw\"ock}, \citenamefont {Silva},
  \citenamefont {Surer}, \citenamefont {Todo}, \citenamefont {Trebst},
  \citenamefont {Troyer}, \citenamefont {Wall}, \citenamefont {Werner},\ and\
  \citenamefont {Wessel}}]{ALPS2}%
  \BibitemOpen
  \bibfield  {author} {\bibinfo {author} {\bibfnamefont {B.}~\bibnamefont
  {Bauer}}, \bibinfo {author} {\bibfnamefont {L.~T.}\ \bibnamefont {Carr}},
  \bibinfo {author} {\bibfnamefont {H.~G.}\ \bibnamefont {Evertz}}, \bibinfo
  {author} {\bibfnamefont {A.~E.}\ \bibnamefont {Feiguin}}, \bibinfo {author}
  {\bibfnamefont {J.}~\bibnamefont {Freire}}, \bibinfo {author} {\bibfnamefont
  {S.}~\bibnamefont {Fuchs}}, \bibinfo {author} {\bibfnamefont
  {L.}~\bibnamefont {Gamper}}, \bibinfo {author} {\bibfnamefont
  {J.}~\bibnamefont {Gukelberger}}, \bibinfo {author} {\bibfnamefont
  {E.}~\bibnamefont {Gull}}, \bibinfo {author} {\bibfnamefont {S.}~\bibnamefont
  {Guertler}}, \bibinfo {author} {\bibfnamefont {A.}~\bibnamefont {Hehn}},
  \bibinfo {author} {\bibfnamefont {R.}~\bibnamefont {Igarashi}}, \bibinfo
  {author} {\bibfnamefont {S.~V.}\ \bibnamefont {Isakov}}, \bibinfo {author}
  {\bibfnamefont {D.}~\bibnamefont {Koop}}, \bibinfo {author} {\bibfnamefont
  {P.~N.}\ \bibnamefont {Ma}}, \bibinfo {author} {\bibfnamefont
  {P.}~\bibnamefont {Mates}}, \bibinfo {author} {\bibfnamefont
  {H.}~\bibnamefont {Matsuo}}, \bibinfo {author} {\bibfnamefont
  {O.}~\bibnamefont {Parcollet}}, \bibinfo {author} {\bibfnamefont
  {G.}~\bibnamefont {Pawlowski}}, \bibinfo {author} {\bibfnamefont {J.~D.}\
  \bibnamefont {Picon}}, \bibinfo {author} {\bibfnamefont {L.}~\bibnamefont
  {Pollet}}, \bibinfo {author} {\bibfnamefont {E.}~\bibnamefont {Santos}},
  \bibinfo {author} {\bibfnamefont {V.~W.}\ \bibnamefont {Scarola}}, \bibinfo
  {author} {\bibfnamefont {U.}~\bibnamefont {Schollw\"ock}}, \bibinfo {author}
  {\bibfnamefont {C.}~\bibnamefont {Silva}}, \bibinfo {author} {\bibfnamefont
  {B.}~\bibnamefont {Surer}}, \bibinfo {author} {\bibfnamefont
  {S.}~\bibnamefont {Todo}}, \bibinfo {author} {\bibfnamefont {S.}~\bibnamefont
  {Trebst}}, \bibinfo {author} {\bibfnamefont {M.}~\bibnamefont {Troyer}},
  \bibinfo {author} {\bibfnamefont {M.~L.}\ \bibnamefont {Wall}}, \bibinfo
  {author} {\bibfnamefont {P.}~\bibnamefont {Werner}},\ and\ \bibinfo {author}
  {\bibfnamefont {S.}~\bibnamefont {Wessel}},\ }\bibfield  {title} {\bibinfo
  {title} {{The ALPS project release 2.0: open source softsoft for strongly
  correlated Systems}},\ }\href
  {https://doi.org/10.1088/1742-5468/2011/05/p05001} {\bibfield  {journal}
  {\bibinfo  {journal} {Journal of Statistical Mechanics: Theory and
  Experiment}\ }\textbf {\bibinfo {volume} {2011}},\ \bibinfo {pages} {P05001}
  (\bibinfo {year} {2011})}\BibitemShut {NoStop}%
\bibitem [{\citenamefont {Ceperley}\ and\ \citenamefont
  {Pollock}(1989)}]{Ceperley1989}%
  \BibitemOpen
  \bibfield  {author} {\bibinfo {author} {\bibfnamefont {D.~M.}\ \bibnamefont
  {Ceperley}}\ and\ \bibinfo {author} {\bibfnamefont {E.~L.}\ \bibnamefont
  {Pollock}},\ }\bibfield  {title} {\bibinfo {title} {Path-integral simulation
  of the superfluid transition in two-dimensional $^{4}\mathrm{He}$},\ }\href
  {https://doi.org/10.1103/PhysRevB.39.2084} {\bibfield  {journal} {\bibinfo
  {journal} {Phys. Rev. B}\ }\textbf {\bibinfo {volume} {39}},\ \bibinfo
  {pages} {2084} (\bibinfo {year} {1989})}\BibitemShut {NoStop}%
\bibitem [{\citenamefont {Koteswararao}\ \emph {et~al.}(2017)\citenamefont
  {Koteswararao}, \citenamefont {Khuntia}, \citenamefont {Kumar}, \citenamefont
  {Mahajan}, \citenamefont {Yogi}, \citenamefont {Baenitz}, \citenamefont
  {Skourski},\ and\ \citenamefont {Chou}}]{Koti2017}%
  \BibitemOpen
  \bibfield  {author} {\bibinfo {author} {\bibfnamefont {B.}~\bibnamefont
  {Koteswararao}}, \bibinfo {author} {\bibfnamefont {P.}~\bibnamefont
  {Khuntia}}, \bibinfo {author} {\bibfnamefont {R.}~\bibnamefont {Kumar}},
  \bibinfo {author} {\bibfnamefont {A.~V.}\ \bibnamefont {Mahajan}}, \bibinfo
  {author} {\bibfnamefont {A.}~\bibnamefont {Yogi}}, \bibinfo {author}
  {\bibfnamefont {M.}~\bibnamefont {Baenitz}}, \bibinfo {author} {\bibfnamefont
  {Y.}~\bibnamefont {Skourski}},\ and\ \bibinfo {author} {\bibfnamefont
  {F.~C.}\ \bibnamefont {Chou}},\ }\bibfield  {title} {\bibinfo {title}
  {{Bose-Einstein condensation of triplons in the $S=1$ tetramer
  antiferromagnet
  ${\mathrm{K}}_{2}{\mathrm{Ni}}_{2}{({\mathrm{MoO}}_{4})}_{3}$: A compound
  close to a quantum critical point}},\ }\href
  {https://doi.org/10.1103/PhysRevB.95.180407} {\bibfield  {journal} {\bibinfo
  {journal} {Phys. Rev. B}\ }\textbf {\bibinfo {volume} {95}},\ \bibinfo
  {pages} {180407} (\bibinfo {year} {2017})}\BibitemShut {NoStop}%
\bibitem [{\citenamefont {Momma}\ and\ \citenamefont {Izumi}(2011)}]{VESTA}%
  \BibitemOpen
  \bibfield  {author} {\bibinfo {author} {\bibfnamefont {K.}~\bibnamefont
  {Momma}}\ and\ \bibinfo {author} {\bibfnamefont {F.}~\bibnamefont {Izumi}},\
  }\bibfield  {title} {\bibinfo {title} {{{\it VESTA3} for three-dimensional
  visualization of crystal, volumetric and morphology data}},\ }\href
  {https://doi.org/10.1107/S0021889811038970} {\bibfield  {journal} {\bibinfo
  {journal} {Journal of Applied Crystallography}\ }\textbf {\bibinfo {volume}
  {44}},\ \bibinfo {pages} {1272} (\bibinfo {year} {2011})}\BibitemShut
  {NoStop}%
\bibitem [{\citenamefont {L{\"o}wdin}(1951)}]{Loewdin1951}%
  \BibitemOpen
  \bibfield  {author} {\bibinfo {author} {\bibfnamefont {P.-O.}\ \bibnamefont
  {L{\"o}wdin}},\ }\bibfield  {title} {\bibinfo {title} {{A Note on the
  Quantum-Mechanical Perturbation Theory}},\ }\href
  {https://doi.org/10.1063/1.1748067} {\bibfield  {journal} {\bibinfo
  {journal} {The Journal of Chemical Physics}\ }\textbf {\bibinfo {volume}
  {19}},\ \bibinfo {pages} {1396} (\bibinfo {year} {1951})}\BibitemShut
  {NoStop}%
\bibitem [{\citenamefont {Wills}\ \emph {et~al.}(2000)\citenamefont {Wills},
  \citenamefont {Eriksson}, \citenamefont {Alouni},\ and\ \citenamefont
  {Price}}]{FPLMTOCode}%
  \BibitemOpen
  \bibfield  {author} {\bibinfo {author} {\bibfnamefont {J.~M.}\ \bibnamefont
  {Wills}}, \bibinfo {author} {\bibfnamefont {O.}~\bibnamefont {Eriksson}},
  \bibinfo {author} {\bibfnamefont {M.}~\bibnamefont {Alouni}},\ and\ \bibinfo
  {author} {\bibfnamefont {D.~L.}\ \bibnamefont {Price}},\ }\href@noop {}
  {\emph {\bibinfo {title} {{Electronic Structure and Physical Properties of
  Solids: The Uses of the LMTO Method}}}}\ (\bibinfo  {publisher}
  {Springer-Verlag},\ \bibinfo {address} {Berlin},\ \bibinfo {year}
  {2000})\BibitemShut {NoStop}%
\bibitem [{\citenamefont {Kvashnin}\ \emph {et~al.}(2015)\citenamefont
  {Kvashnin}, \citenamefont {Gr\aa{}n\"as}, \citenamefont {Di~Marco},
  \citenamefont {Katsnelson}, \citenamefont {Lichtenstein},\ and\ \citenamefont
  {Eriksson}}]{Exchange_Rspt}%
  \BibitemOpen
  \bibfield  {author} {\bibinfo {author} {\bibfnamefont {Y.~O.}\ \bibnamefont
  {Kvashnin}}, \bibinfo {author} {\bibfnamefont {O.}~\bibnamefont
  {Gr\aa{}n\"as}}, \bibinfo {author} {\bibfnamefont {I.}~\bibnamefont
  {Di~Marco}}, \bibinfo {author} {\bibfnamefont {M.~I.}\ \bibnamefont
  {Katsnelson}}, \bibinfo {author} {\bibfnamefont {A.~I.}\ \bibnamefont
  {Lichtenstein}},\ and\ \bibinfo {author} {\bibfnamefont {O.}~\bibnamefont
  {Eriksson}},\ }\bibfield  {title} {\bibinfo {title} {{Exchange parameters of
  strongly correlated materials: Extraction from spin-polarized density
  functional theory plus dynamical mean-field theory}},\ }\href
  {https://doi.org/10.1103/PhysRevB.91.125133} {\bibfield  {journal} {\bibinfo
  {journal} {Phys. Rev. B}\ }\textbf {\bibinfo {volume} {91}},\ \bibinfo
  {pages} {125133} (\bibinfo {year} {2015})}\BibitemShut {NoStop}%
\bibitem [{\citenamefont {Liechtenstein}\ \emph {et~al.}(1995)\citenamefont
  {Liechtenstein}, \citenamefont {Anisimov},\ and\ \citenamefont
  {Zaanen}}]{Liechtenstein1995}%
  \BibitemOpen
  \bibfield  {author} {\bibinfo {author} {\bibfnamefont {A.~I.}\ \bibnamefont
  {Liechtenstein}}, \bibinfo {author} {\bibfnamefont {V.~I.}\ \bibnamefont
  {Anisimov}},\ and\ \bibinfo {author} {\bibfnamefont {J.}~\bibnamefont
  {Zaanen}},\ }\bibfield  {title} {\bibinfo {title} {{Density-functional theory
  and strong interactions: Orbital ordering in Mott-Hubbard insulators}},\
  }\href {https://doi.org/10.1103/PhysRevB.52.R5467} {\bibfield  {journal}
  {\bibinfo  {journal} {Phys. Rev. B}\ }\textbf {\bibinfo {volume} {52}},\
  \bibinfo {pages} {R5467} (\bibinfo {year} {1995})}\BibitemShut {NoStop}%
\bibitem [{\citenamefont {Bl\"ochl}(1994)}]{Blochl1994}%
  \BibitemOpen
  \bibfield  {author} {\bibinfo {author} {\bibfnamefont {P.~E.}\ \bibnamefont
  {Bl\"ochl}},\ }\bibfield  {title} {\bibinfo {title} {{Projector
  augmented-wave method}},\ }\href {https://doi.org/10.1103/PhysRevB.50.17953}
  {\bibfield  {journal} {\bibinfo  {journal} {Phys. Rev. B}\ }\textbf {\bibinfo
  {volume} {50}},\ \bibinfo {pages} {17953} (\bibinfo {year}
  {1994})}\BibitemShut {NoStop}%
\bibitem [{\citenamefont {Kresse}\ and\ \citenamefont
  {Furthm\"uller}(1996)}]{Kresse1996}%
  \BibitemOpen
  \bibfield  {author} {\bibinfo {author} {\bibfnamefont {G.}~\bibnamefont
  {Kresse}}\ and\ \bibinfo {author} {\bibfnamefont {J.}~\bibnamefont
  {Furthm\"uller}},\ }\bibfield  {title} {\bibinfo {title} {{Efficient
  iterative schemes for ab initio total-energy calculations using a plane-wave
  basis set}},\ }\href {https://doi.org/10.1103/PhysRevB.54.11169} {\bibfield
  {journal} {\bibinfo  {journal} {Phys. Rev. B}\ }\textbf {\bibinfo {volume}
  {54}},\ \bibinfo {pages} {11169} (\bibinfo {year} {1996})}\BibitemShut
  {NoStop}%
\bibitem [{\citenamefont {Blaha}\ \emph {et~al.}(1990)\citenamefont {Blaha},
  \citenamefont {Schwarz}, \citenamefont {Sorantin},\ and\ \citenamefont
  {Trickey}}]{Blaha1999}%
  \BibitemOpen
  \bibfield  {author} {\bibinfo {author} {\bibfnamefont {P.}~\bibnamefont
  {Blaha}}, \bibinfo {author} {\bibfnamefont {K.}~\bibnamefont {Schwarz}},
  \bibinfo {author} {\bibfnamefont {P.}~\bibnamefont {Sorantin}},\ and\
  \bibinfo {author} {\bibfnamefont {S.}~\bibnamefont {Trickey}},\ }\bibfield
  {title} {\bibinfo {title} {{Full-potential, linearized augmented plane wave
  programs for crystalline systems}},\ }\href
  {https://doi.org/https://doi.org/10.1016/0010-4655(90)90187-6} {\bibfield
  {journal} {\bibinfo  {journal} {Computer Physics Communications}\ }\textbf
  {\bibinfo {volume} {59}},\ \bibinfo {pages} {399 } (\bibinfo {year}
  {1990})}\BibitemShut {NoStop}%
\bibitem [{\citenamefont {Wills}\ and\ \citenamefont {Cooper}(1987)}]{FPLMTO}%
  \BibitemOpen
  \bibfield  {author} {\bibinfo {author} {\bibfnamefont {J.~M.}\ \bibnamefont
  {Wills}}\ and\ \bibinfo {author} {\bibfnamefont {B.~R.}\ \bibnamefont
  {Cooper}},\ }\bibfield  {title} {\bibinfo {title} {{Synthesis of band and
  model Hamiltonian theory for hybridizing cerium systems}},\ }\href
  {https://doi.org/10.1103/PhysRevB.36.3809} {\bibfield  {journal} {\bibinfo
  {journal} {Phys. Rev. B}\ }\textbf {\bibinfo {volume} {36}},\ \bibinfo
  {pages} {3809} (\bibinfo {year} {1987})}\BibitemShut {NoStop}%
\bibitem [{\citenamefont {Andersen}(1975)}]{FPLMTO_Orig}%
  \BibitemOpen
  \bibfield  {author} {\bibinfo {author} {\bibfnamefont {O.~K.}\ \bibnamefont
  {Andersen}},\ }\bibfield  {title} {\bibinfo {title} {{Linear methods in band
  theory}},\ }\href {https://doi.org/10.1103/PhysRevB.12.3060} {\bibfield
  {journal} {\bibinfo  {journal} {Phys. Rev. B}\ }\textbf {\bibinfo {volume}
  {12}},\ \bibinfo {pages} {3060} (\bibinfo {year} {1975})}\BibitemShut
  {NoStop}%
\bibitem [{\citenamefont {Panda}\ \emph {et~al.}(2017)\citenamefont {Panda},
  \citenamefont {Jiang},\ and\ \citenamefont {Biermann}}]{PhysRevB.96.045137}%
  \BibitemOpen
  \bibfield  {author} {\bibinfo {author} {\bibfnamefont {S.~K.}\ \bibnamefont
  {Panda}}, \bibinfo {author} {\bibfnamefont {H.}~\bibnamefont {Jiang}},\ and\
  \bibinfo {author} {\bibfnamefont {S.}~\bibnamefont {Biermann}},\ }\bibfield
  {title} {\bibinfo {title} {{Pressure dependence of dynamically screened
  Coulomb interactions in NiO: Effective Hubbard, Hund, intershell, and
  intersite components}},\ }\href {https://doi.org/10.1103/PhysRevB.96.045137}
  {\bibfield  {journal} {\bibinfo  {journal} {Phys. Rev. B}\ }\textbf {\bibinfo
  {volume} {96}},\ \bibinfo {pages} {045137} (\bibinfo {year}
  {2017})}\BibitemShut {NoStop}%
\bibitem [{\citenamefont {Rodríguez-Carvajal}(1993)}]{fullprof}%
  \BibitemOpen
  \bibfield  {author} {\bibinfo {author} {\bibfnamefont {J.}~\bibnamefont
  {Rodríguez-Carvajal}},\ }\bibfield  {title} {\bibinfo {title} {{Recent
  advances in magnetic structure determination by neutron powder
  diffraction}},\ }\href
  {https://doi.org/https://doi.org/10.1016/0921-4526(93)90108-I} {\bibfield
  {journal} {\bibinfo  {journal} {Physica B: Condensed Matter}\ }\textbf
  {\bibinfo {volume} {192}},\ \bibinfo {pages} {55 } (\bibinfo {year}
  {1993})}\BibitemShut {NoStop}%
\bibitem [{\citenamefont {Mostofi}\ \emph {et~al.}(2008)\citenamefont
  {Mostofi}, \citenamefont {Yates}, \citenamefont {Lee}, \citenamefont {Souza},
  \citenamefont {Vanderbilt},\ and\ \citenamefont {Marzari}}]{wannier90}%
  \BibitemOpen
  \bibfield  {author} {\bibinfo {author} {\bibfnamefont {A.~A.}\ \bibnamefont
  {Mostofi}}, \bibinfo {author} {\bibfnamefont {J.~R.}\ \bibnamefont {Yates}},
  \bibinfo {author} {\bibfnamefont {Y.-S.}\ \bibnamefont {Lee}}, \bibinfo
  {author} {\bibfnamefont {I.}~\bibnamefont {Souza}}, \bibinfo {author}
  {\bibfnamefont {D.}~\bibnamefont {Vanderbilt}},\ and\ \bibinfo {author}
  {\bibfnamefont {N.}~\bibnamefont {Marzari}},\ }\bibfield  {title} {\bibinfo
  {title} {{wannier90: A tool for obtaining maximally-localised Wannier
  functions}},\ }\href {https://doi.org/10.1016/j.cpc.2007.11.016} {\bibfield
  {journal} {\bibinfo  {journal} {Computer Physics Communications}\ }\textbf
  {\bibinfo {volume} {178}},\ \bibinfo {pages} {685 } (\bibinfo {year}
  {2008})}\BibitemShut {NoStop}%
\bibitem [{\citenamefont {Kuneš}\ \emph {et~al.}(2010)\citenamefont {Kuneš},
  \citenamefont {Arita}, \citenamefont {Wissgott}, \citenamefont {Toschi},
  \citenamefont {Ikeda},\ and\ \citenamefont {Held}}]{Wien2wannier}%
  \BibitemOpen
  \bibfield  {author} {\bibinfo {author} {\bibfnamefont {J.}~\bibnamefont
  {Kuneš}}, \bibinfo {author} {\bibfnamefont {R.}~\bibnamefont {Arita}},
  \bibinfo {author} {\bibfnamefont {P.}~\bibnamefont {Wissgott}}, \bibinfo
  {author} {\bibfnamefont {A.}~\bibnamefont {Toschi}}, \bibinfo {author}
  {\bibfnamefont {H.}~\bibnamefont {Ikeda}},\ and\ \bibinfo {author}
  {\bibfnamefont {K.}~\bibnamefont {Held}},\ }\bibfield  {title} {\bibinfo
  {title} {{Wien2wannier: From linearized augmented plane waves to maximally
  localized Wannier functions}},\ }\href
  {https://doi.org/10.1016/j.cpc.2010.08.005} {\bibfield  {journal} {\bibinfo
  {journal} {Computer Physics Communications}\ }\textbf {\bibinfo {volume}
  {181}},\ \bibinfo {pages} {1888 } (\bibinfo {year} {2010})}\BibitemShut
  {NoStop}%
\bibitem [{\citenamefont {Boninsegni}\ and\ \citenamefont
  {Prokof'ev}(2005)}]{Boninsegni2005}%
  \BibitemOpen
  \bibfield  {author} {\bibinfo {author} {\bibfnamefont {M.}~\bibnamefont
  {Boninsegni}}\ and\ \bibinfo {author} {\bibfnamefont {N.}~\bibnamefont
  {Prokof'ev}},\ }\bibfield  {title} {\bibinfo {title} {{Supersolid Phase of
  Hard-Core Bosons on a Triangular Lattice}},\ }\href
  {https://doi.org/10.1103/PhysRevLett.95.237204} {\bibfield  {journal}
  {\bibinfo  {journal} {Phys. Rev. Lett.}\ }\textbf {\bibinfo {volume} {95}},\
  \bibinfo {pages} {237204} (\bibinfo {year} {2005})}\BibitemShut {NoStop}%
\bibitem [{\citenamefont {Wessel}\ and\ \citenamefont
  {Troyer}(2005)}]{Wessel2005}%
  \BibitemOpen
  \bibfield  {author} {\bibinfo {author} {\bibfnamefont {S.}~\bibnamefont
  {Wessel}}\ and\ \bibinfo {author} {\bibfnamefont {M.}~\bibnamefont
  {Troyer}},\ }\bibfield  {title} {\bibinfo {title} {{Supersolid Hard-Core
  Bosons on the Triangular Lattice}},\ }\href
  {https://doi.org/10.1103/PhysRevLett.95.127205} {\bibfield  {journal}
  {\bibinfo  {journal} {Phys. Rev. Lett.}\ }\textbf {\bibinfo {volume} {95}},\
  \bibinfo {pages} {127205} (\bibinfo {year} {2005})}\BibitemShut {NoStop}%
\bibitem [{\citenamefont {Heidarian}\ and\ \citenamefont
  {Damle}(2005)}]{Heidarian2005}%
  \BibitemOpen
  \bibfield  {author} {\bibinfo {author} {\bibfnamefont {D.}~\bibnamefont
  {Heidarian}}\ and\ \bibinfo {author} {\bibfnamefont {K.}~\bibnamefont
  {Damle}},\ }\bibfield  {title} {\bibinfo {title} {{Persistent Supersolid
  Phase of Hard-Core Bosons on the Triangular Lattice}},\ }\href
  {https://doi.org/10.1103/PhysRevLett.95.127206} {\bibfield  {journal}
  {\bibinfo  {journal} {Phys. Rev. Lett.}\ }\textbf {\bibinfo {volume} {95}},\
  \bibinfo {pages} {127206} (\bibinfo {year} {2005})}\BibitemShut {NoStop}%
\bibitem [{\citenamefont {Melko}\ \emph {et~al.}(2005)\citenamefont {Melko},
  \citenamefont {Paramekanti}, \citenamefont {Burkov}, \citenamefont
  {Vishwanath}, \citenamefont {Sheng},\ and\ \citenamefont
  {Balents}}]{Melko2005}%
  \BibitemOpen
  \bibfield  {author} {\bibinfo {author} {\bibfnamefont {R.~G.}\ \bibnamefont
  {Melko}}, \bibinfo {author} {\bibfnamefont {A.}~\bibnamefont {Paramekanti}},
  \bibinfo {author} {\bibfnamefont {A.~A.}\ \bibnamefont {Burkov}}, \bibinfo
  {author} {\bibfnamefont {A.}~\bibnamefont {Vishwanath}}, \bibinfo {author}
  {\bibfnamefont {D.~N.}\ \bibnamefont {Sheng}},\ and\ \bibinfo {author}
  {\bibfnamefont {L.}~\bibnamefont {Balents}},\ }\bibfield  {title} {\bibinfo
  {title} {{Supersolid Order from Disorder: Hard-Core Bosons on the Triangular
  Lattice}},\ }\href {https://doi.org/10.1103/PhysRevLett.95.127207} {\bibfield
   {journal} {\bibinfo  {journal} {Phys. Rev. Lett.}\ }\textbf {\bibinfo
  {volume} {95}},\ \bibinfo {pages} {127207} (\bibinfo {year}
  {2005})}\BibitemShut {NoStop}%
\bibitem [{\citenamefont {Gan}(2008)}]{Gan2008}%
  \BibitemOpen
  \bibfield  {author} {\bibinfo {author} {\bibfnamefont {J.-Y.}\ \bibnamefont
  {Gan}},\ }\bibfield  {title} {\bibinfo {title} {{Effects of frustration on
  the anisotropic triangular lattice bosons}},\ }\href
  {https://doi.org/10.1103/PhysRevB.78.014513} {\bibfield  {journal} {\bibinfo
  {journal} {Phys. Rev. B}\ }\textbf {\bibinfo {volume} {78}},\ \bibinfo
  {pages} {014513} (\bibinfo {year} {2008})}\BibitemShut {NoStop}%
\bibitem [{\citenamefont {Zhang}\ \emph {et~al.}(2016)\citenamefont {Zhang},
  \citenamefont {Hu}, \citenamefont {Pelster},\ and\ \citenamefont
  {Eggert}}]{Zhang2016}%
  \BibitemOpen
  \bibfield  {author} {\bibinfo {author} {\bibfnamefont {X.-F.}\ \bibnamefont
  {Zhang}}, \bibinfo {author} {\bibfnamefont {S.}~\bibnamefont {Hu}}, \bibinfo
  {author} {\bibfnamefont {A.}~\bibnamefont {Pelster}},\ and\ \bibinfo {author}
  {\bibfnamefont {S.}~\bibnamefont {Eggert}},\ }\bibfield  {title} {\bibinfo
  {title} {{Quantum Domain Walls Induce Incommensurate Supersolid Phase on the
  Anisotropic Triangular Lattice}},\ }\href
  {https://doi.org/10.1103/PhysRevLett.117.193201} {\bibfield  {journal}
  {\bibinfo  {journal} {Phys. Rev. Lett.}\ }\textbf {\bibinfo {volume} {117}},\
  \bibinfo {pages} {193201} (\bibinfo {year} {2016})}\BibitemShut {NoStop}%
\bibitem [{\citenamefont {Zvyagin}\ \emph {et~al.}(2006)\citenamefont
  {Zvyagin}, \citenamefont {Wosnitza}, \citenamefont {Krzystek}, \citenamefont
  {Stern}, \citenamefont {Jaime}, \citenamefont {Sasago},\ and\ \citenamefont
  {Uchinokura}}]{Zvyagin2006}%
  \BibitemOpen
  \bibfield  {author} {\bibinfo {author} {\bibfnamefont {S.~A.}\ \bibnamefont
  {Zvyagin}}, \bibinfo {author} {\bibfnamefont {J.}~\bibnamefont {Wosnitza}},
  \bibinfo {author} {\bibfnamefont {J.}~\bibnamefont {Krzystek}}, \bibinfo
  {author} {\bibfnamefont {R.}~\bibnamefont {Stern}}, \bibinfo {author}
  {\bibfnamefont {M.}~\bibnamefont {Jaime}}, \bibinfo {author} {\bibfnamefont
  {Y.}~\bibnamefont {Sasago}},\ and\ \bibinfo {author} {\bibfnamefont
  {K.}~\bibnamefont {Uchinokura}},\ }\bibfield  {title} {\bibinfo {title}
  {{Spin-triplet excitons in the $S=\frac{1}{2}$ gapped antiferromagnet
  $\mathrm{Ba}\mathrm{Cu}{\mathrm{Si}}_{2}{\mathrm{O}}_{6}$: Electron
  paramagnetic resonance studies}},\ }\href
  {https://doi.org/10.1103/PhysRevB.73.094446} {\bibfield  {journal} {\bibinfo
  {journal} {Phys. Rev. B}\ }\textbf {\bibinfo {volume} {73}},\ \bibinfo
  {pages} {094446} (\bibinfo {year} {2006})}\BibitemShut {NoStop}%
\end{thebibliography}
\end{document}